\documentclass[11pt,journal,onecolumn,letterpaper]{IEEEtran}

\usepackage{listings}
\usepackage{float}
\usepackage{setspace}
\usepackage{amsmath}

\usepackage{makecell}
\usepackage{mathabx}
\usepackage{array}
\newcolumntype{L}[1]{>{\raggedright\arraybackslash} m{#1} }
\usepackage{subcaption}
\usepackage{graphicx}  
\usepackage{epstopdf}
\usepackage{mdwmath}
\usepackage{tabu}
\usepackage{mdwtab}
\usepackage{float}
\usepackage{filecontents}
\usepackage{eqparbox}
\usepackage{url}
\usepackage[english]{babel}
\usepackage[utf8]{inputenc}
\usepackage{cite}
\usepackage{algorithm}
\usepackage{tikz}
\usepackage{algpseudocode}
\usepackage{varwidth}
\usetikzlibrary{shapes.misc}
\usepackage{mathptmx}

\usepackage[T1]{fontenc}

\usepackage[T1]{fontenc}
\usepackage{geometry}
\usepackage{enumitem}
\usepackage{tikz}
\usetikzlibrary{automata,arrows,positioning,calc}
\usepackage{booktabs}
\usetikzlibrary{backgrounds,calc,shadings,shapes.arrows,shapes.symbols,shadows}
\definecolor{switch}{HTML}{006996}
\algrenewcommand\textproc{}
\usepackage{multicol}
\usepackage{multirow}
\newcolumntype{P}[1]{>{\centering\arraybackslash}p{#1}}
\begin{document}

\title{
\textbf{TECHNICAL REPORT} \\ 
Design and Implementation of SMARTHO -- A Network Initiated Handover
mechanism in NG-RAN, on  P4-based Xilinx NetFPGA switches
}
\author{\IEEEauthorblockN{\textsf{Phanindra Palagummi and Krishna
      M. Sivalingam}} \\
\IEEEauthorblockA{\textit{Department of Computer Science and
    Engineering \\ Indian Institute of Technology
    Madras, Chennai, INDIA} \\
 Emails: \{phanindra.palagummi@gmail.com, skrishnam@iitm.ac.in,
 krishna.sivalingam@gmail.com\}} \\
\today
}
\maketitle
\begin{abstract}

This report deals with the design of handover schemes for radio access
networks (RAN) in 5G networks, using programmable data plane
switches. The network architecture is expected to be a centralized
cloud infrastructure, connected via a backhaul network to many
edge-computing clouds that are closer to the end-user. Some of the
network services can be implemented in edge devices to improve network
performance.

In 5G networks, the C-RAN architecture splits the Base Band Unit (BBU)
into Central and Distributed Units (CU and DU). This structure has
created a mid-haul Network, connecting CUs and DUs. The mid-haul
network has created a dataplane challenge that does not exist in
traditional distributed RANs -- the need for efficient connections
between the CUs and DUs. Traditional encapsulation techniques can be
used to transport packets across the CU and DU. However, the recent
advancements in dataplane programmability can be used to enhance the
system performance. In this report, we show how P4 switches can be used
to parse the packets between DU, CU, and Back Haul (Core Network) for
potential system improvements. In particular, we consider the scenario
of mobile handover, that arises when a user moves between different
cells in the mobile network. The proposed protocol is called
\textit{SMARTHO}, illustrating a smart handover.

\textit{Programming Protocol-Independent Packet Processors (P4)} is a
programming language designed to support specification and programming
the forwarding plane behavior of network switches/routers.  With P4
switches, the protocol designer can define customized packet headers,
parsing of headers, and defining new match-action routines. In
\textit{SMARTHO}, we use P4 Switches to intervene in the handover
process for fixed-path mobile users. Such users could be those in a
train, drones, devices with high-degree of predictable mobility,
etc. A resource pre-allocation scheme that reserves resources before
the UE reaches a future cell, is proposed. The solution is implemented
using a P4-based switch introduced between the CU and the DU.  The P4
switch is used to spoof the behavior of User Equipment (UE) and
perform the resource allocation in advance. This is expected to reduce
the handover time as the user moves along its path.

The proposed SMARTHO framework is implemented in the \textit{mininet}
emulation environment and in a reconfigurable hardware environment
using NetFPGA-SUME boards. For Mininet based simulation, we used
virtual hosts connected using P4 switches, using the P4 behavior model
(P4BMv2) software switch. User and control traffic is also generated
to simulate the mobile traffic and measure the HO performance. User
traffic is represented using ICMP ping packets over a tag. The results
show a handover response time improvement of 18\% for a tandem of two
HOs and 25\% for a tandem of three HOs. For testbed implementation, we
used NetFPGA-SUME boards as P4 switches. The Xilinx SDNet tool-chain
is used to compile P4 programs directly to NetFPGA-SUME. Raw data
packets are generated using the \textit{scapy} tool. The handover time
was measured to be approximately 50~milliseconds in the experiments
conducted.  
\end{abstract}


\section{Introduction}

This report deals with improving handover performance in 5G Wireless
networks, using the programmable data plane switch paradigm.  A large
number of operators are now evaluating Next-Generation RAN (NG-RAN) as
a way to meet future service requirements.  NG-RAN is an enhancement
to the earlier Cloud-RAN (C-RAN) architecture that is
fully-centralized and fixed, but not adaptive to network traffic.
Part of this work was published as a short paper \cite{CNSM18} and as a
M.S. (by Research) Thesis at Indian Institute of Technology Madras,
Chennai, INDIA \cite{PhaniThesis}.

In the NG-RAN architecture, real-time (RT) functions are deployed near
the antenna site to manage air interface resources, by the Distributed
Units (DU).  At the same time, non-real-time (NRT) control functions
are hosted centrally in the Central Unit.  This split functionality is
now part of the 3GPP specification\cite{3gpp38401}.  The services
offered by the CU and DU can be virtualized in software and placed in
Commercial off-the-shelf (COTS) servers, using Network Function
Virtualizaton (NFV)
\cite{giannoulakis2014applications,hawilo2014nfv,abdelwahab2016network,costa2015sdn}.

In this report, we design a solution for handling mobile device
handover, using programmable data-plane switches based on P4
programming language\cite{bosshart2014p4}.  P4-based switches are used
to parse the packets and to invoke additional actions defined by the
protocol designer.  These actions can be made to perform simple
forwarding or can aid functional behaviour of the system.

\begin{figure}[hbtp]
\centering
  \includegraphics[width=0.7\linewidth]{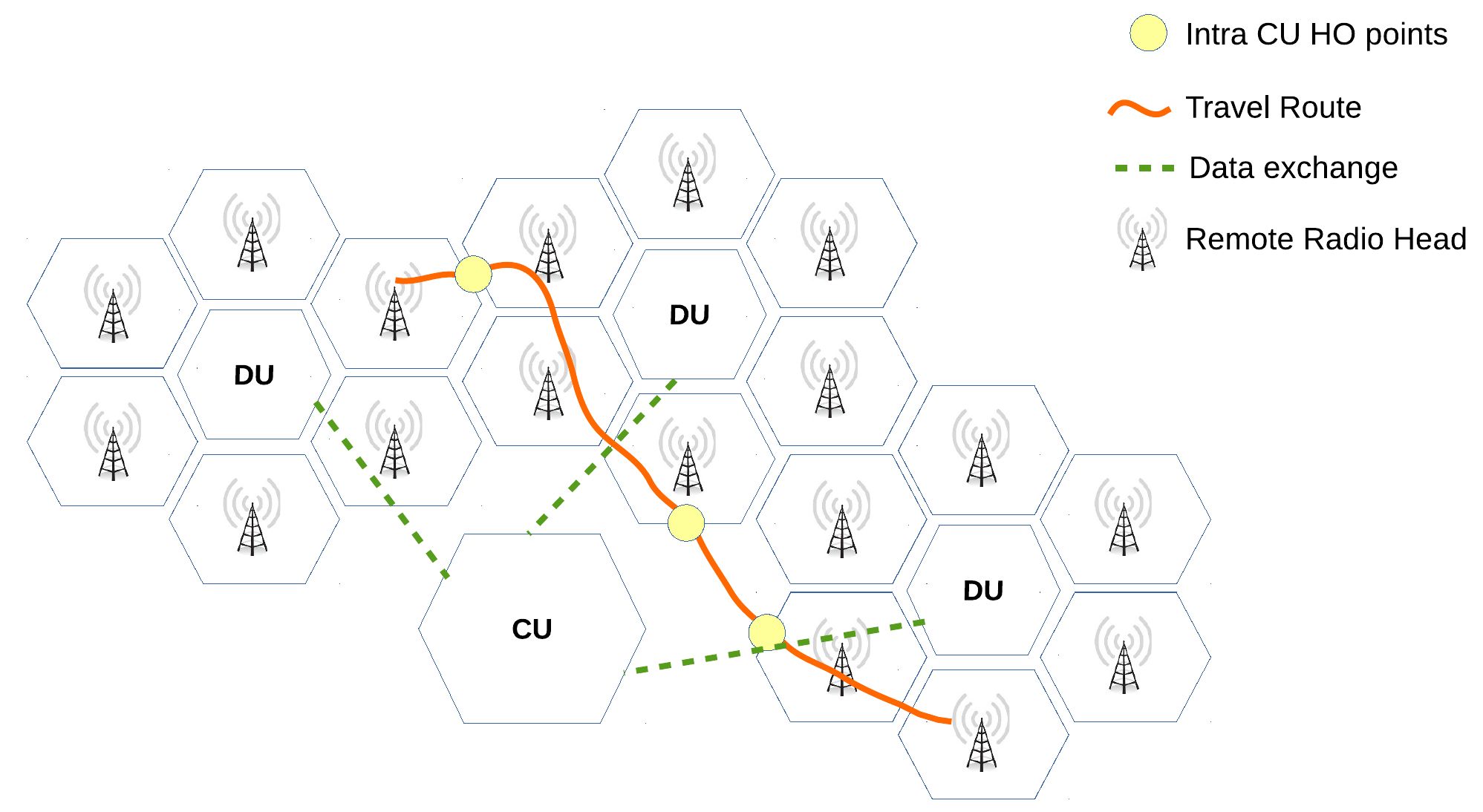}
    \caption{Intra-CU Handover.}\label{5gintracuho1}
\end{figure}

In particular, we propose a Smart Handover (SMARTHO) process for
fixed-path mobile devices, such as LTE users in a train, drones,
predictable mobility devices, etc. is considered. In particular, the
handover is considered for Intra-CU HO from one Radio Head (RH) to
another RH in a different DU, but connected to the same CU. This
scenario is shown in Figure~\ref{5gintracuho1}.  A resource allocation
scheme that reserves resources ahead of the UE in its path is
proposed. The solution is implemented using a P4-based switch
introduced between the CU and the DU.  We use the P4 switch to spoof
the behaviour of User Equipment (UE) and perform the resource
allocation in advance.  Using an implementation based on Mininet and
P4BM software switch, it is seen that the proposed method results in
an 18\% and 25\% improvement in the sequence of two and three
handovers, respectively.  A prototype of the mechanism has also been
implemented in a reconfigurable hardware environment using Xilinx
NetFPGA-SUME boards, using the P4 Programmable Data Plane (PDP)
language \cite{bosshart2014p4,netfpgap4,p4}.

We have considered the Intra-CU handover in this report; however, this
idea can be applied to other HO processes specified in 3GPP
\cite{3gpp38401}.

\section{Background} \label{backgr}

This section presents the relevant background material.

\subsection{5G NG-RAN}

There are several service dimensions in 5G networks \cite{3gpp22891},
including support for massive Machine-Type Communications (mMTC),
enhanced Mobile Broadband (eMBB), and Ultra-Reliable Low-Latency
Communications (UR-LCC) services. Each service has very different
performance requirements and traffic profiles. To serve these new
markets and to increase revenues substantially, operators need highly
scalable and flexible networks.  A large number of operators are now
evaluating Next-Generation RAN (NG-RAN) as a way to meet future
service requirements. From the initial days of deploying
Cloud-RAN\cite{checko2015cloud}, which was business oriented to save
operational costs, the focus has now evolved to meet the future
complex and varied service requirements.

\subsubsection{C-RAN and NG-RAN}
The traditional C-RAN architecture is fully-centralized and fixed,
which is not adaptive to the movable traffic and the advanced
software defined networking concepts. As a result, it is urgent to
improve the friable capability of C-RANs. This led the research
community to work on functional split options in C-RAN. In
FluidNet\cite{fluidnet}, the novel concept of re-configurable
fronthaul is proposed, to flexibly support one-to-one and one-to-many
logical mappings between Base Band Units (BBUs) and Radio Resource
Heads (RRHs) to perform proper transmission
strategies. R-FFT\cite{thyagaturu2018r} proposed IFFT/FFT the PHY
layer split, which would reduce the fronthaul bitrate requirements and
enable statistical multiplexing. An optimal functional split is
discussed by wang et al., team\cite{wang2017interplay}. The technical
report\cite{3gpp38801} sets out various options for the RAN and its
interfaces to the core network. 

In the NG-RAN architecture, real-time (RT) functions are deployed near
the antenna site to manage air interface resources, while
non-real-time (NRT) control functions are hosted centrally to
coordinate transmissions across the coverage area. In NG-RAN, this is
being formalized with the Centralized Unit (CU) and Distributed Unit
(DU) functional split. This functional architecture is now native to
the 3GPP specification\cite{3gpp38401}.

\subsection{Architectural Principles of CU and DU Split} \label{archprin}

The implementation of the NG-RAN architecture and its subsequent
deployment in the network depends on the functional split between
distributed radio and centralized control, called the DU-CU split. The
DU will process low-level radio protocol and real-time services while
the CU will process non-real-time radio protocols.

3GPP has recognized eight different split options\cite{3gpp38801}. Of
these option-2 and option-3, are the most widely discussed two
splits. In option-2 the function split will have ``Radio Resources
Control'' (RRC), ``Packet Data Convergence Protocol'' (PDCP) in the
CU. DU will perform the low-level stack of ``Radio Link Control''
(RLC), ``Media Access Control'' (MAC), while the physical layer and RF
will be in Remote Radio Unit (RRU). 

In the option-3 split, low RLC (a partial function of RLC), MAC,
physical layer are in DU. PDCP and high RLC (the other partial
function of RLC) are in the CU. These split options are discussed in
3GPP status meeting\cite{3gppcudusplit}. The services of CU and DU can
be virtualized and put in Commercial off-the-shelf (COTS) servers,
these virtualized network nodes or Virtual Network Functions (VNFs)
can be realized with a network architectural concept called Network
Function Virtualization (NFV)\cite{giannoulakis2014applications}. NFV
offers a new way to design, deploy and manage virtual network
nodes. It also enables us to decouple suppliers hardware and software
business models, opening new innovations and opportunities for SW
integrators.

The management and operational aspects of NG-RAN with CU and DU splits
would be easy to handle using NFV. There are several research papers
which already attempted in virtualizing mobile network functions
\cite{hawilo2014nfv,abdelwahab2016network,costa2015sdn}.

\subsection{Intra-CU Handover}

In a wireless network, user equipment (UE) handover from one cell to
another cell is an important aspect of mobility management. In this
report, we consider intra-DU handover within a single CU.  Typically,
there are 3 phases in a handover (HO) process: Preparation Execution
and Completion.

The preparation phase deals primarily with resource allocation for the
UE in the next DU.  In this phase, the Measurement Report (MR) message
from the Source\_DU will be transmitted to the CU, which would select
the Target\_DU for the HO. The CU will send the HO request (UE Context
Request), containing Target-DU-ID, UE context info \& UE History
Information. When the Target\_DU receives the HO request, it begins
handover preparation to ensure seamless service provision for the
UE. The Target\_DU would respond with setting up Access Stratum (AS)
security keys, uplink bearers connecting to the backhaul, reserve
Radio Resource Control (RRC) resources to be used by the mobile device
over the radio link and allocates Cell-Radio Network Temporary
Identifier. Once the resources are allocated by the Target\_DU, a
response message called the ``UE context setup response'' is sent 
to the CU. 

Once handover preparation between the two DUs (Source\_DU and
Target\_DU) is completed, the execution phase will start to have the
UE perform a handover. The Source\_DU instructs the UE to perform a
handover by sending RRC Connection Reconfiguration message that
includes all the information needed to access the Target\_DU. The
Target\_DU sends an Uplink RRC Transfer message to the CU to convey
the received RRCConnectionReconfigurationComplete message. Then,
downlink packets are sent to the UE. Also, uplink packets are sent
from the UE, which are forwarded to the CU through the Target\_DU.

In the final completion phase, the CU sends the UE context release
command to the Source\_DU which would release all the bearers from CU
to Source\_DU. 

In this report, we deal with the preparation phase, by proposing a
advanced resource allocation scheme along a set of pre-defined DU
nodes. The proposed design, working model and elements involved in
SMARTHO are discussed in Section~\ref{Design}.

\subsection{Related Work}

There are several papers that deal with handover procedures involving
high mobility. We focus on works dealing with handover support for
fixed-path mobile users, such as those on a train.

In \cite{trainseamless}, a dual\_link HO scheme is studied for
wireless Mobile communication in high\_speed rails. Here, an extra
antenna is used, one for handover and other for data communication
with the base station. In \cite{raildistantenna}, a radio-over-fibre
based approach has been proposed to provide communications inside long
tunnels using distributed antenna systems, and performing HO over
these antennae. In \cite{multitunnelmobility}, a multiple-tunnel based
approach with multiple interfaces and a modified "Hierarchical Mobile
IPv6" (HMIPv6) Mobility Management method, is considered. In,
\cite{li2016mobility}, mobility prediction based handover with
RAN-Cache has been studied for HetNets.

In \cite{seamlessHOLTEWIFI}, a variant of Proxy Mobile Internet
Protocol (PMIP) is developed to reduce ping-pong (PP) events and
handover failures. In \cite{mihprotocol}, vertical handover is
considered by introducing a layer between MAC and PHY layers; this
extra layer performs the handover across different technologies. 

A measurement of LTE performance on high velocity environment is
studied in\cite{lteperformanceonltevel}. Some papers have studied
approaches on the Time To Trigger (TTT) for handover. The work in
\cite{mobperfhetnets} showed that a lower value of TTT for HO would
decrease the handover failure, but would increased the ping-pong
effect. The work \cite{handoverinmobility} suggested that handover
margin is more appropriate than TTT to adjust handover timing, in
response to the change in mobility conditions. In
\cite{lterailtriggeropt}, the relation between the TTT and the
position of high speed train was investigated. The work in
\cite{zheng2008performance} presents an integrated HO algorithm in LTE
networks, while a Received Signal Strength (RSS) based TTT algorithm
has been studied in \cite{anas2007performance}.

In all above papers dealing with fixed-path user mobility,
pre-allocation of resources along the path have not been
considered. In this report, We attempt this approach with the use of
programmable data-plane entities.

\subsection{Programmable Data Plane Switches}

The recent Software Defined Networking networking paradigm (SDN) and
associated protocols and implementations such as OpenFlow
Protocol\cite{martinez2015next} and Open VSwitch
(OVS)\cite{pfaff2015design} allow programmability in the data
plane. However, these are are not protocol independent.  When these
switches are used in mobile networks where protocol stack largely
differ from the standard protocols, the forwarding behaviour would be
limited to encapsulation/tunneling mechanisms.  The strict parsers and
forwarding routines can help improve the forwarding
behaviour\cite{hommes2017optimising,macdavid2017concise,chourasia2015sdn},
but would not aid in adding new system functions.

Programming Protocol independent Packet Parsers (P4) provides is an
upcoming framework for realizing programmable data-plane switches
\cite{bosshart2014p4}.  P4 switches are expected to perform better
than traditional L2-L3/ Open Flow switches due to the additional
functionality enabled. For instance, we show that a simple tag based
forwarding approach over an IP-based encapsulation mechanism is
showing 27\% improvement using a P4 behaviour model (P4BM) software
switch.  Hence, this report considered the use of P4-based switches for
improving handover performance in future wireless networks.

\section{Proposed SMARTHO Framework} \label{Design}

This section presents the details of the proposed Smart Handover
(SMARTHO) mobility management framework.

\subsection{SMARTHO Architecture and Components}

This section presents the architecture, components and message
exchanges involved in SMARTHO model. 3GPP has already discussed the
NG-RAN architecture\cite{3gpp38401}. For SMARTHO, we introduce
programmability into the data plane without changing the existing
architectural framework.

\begin{figure}[htbp]
  \centering
  \includegraphics[width=0.6\linewidth]{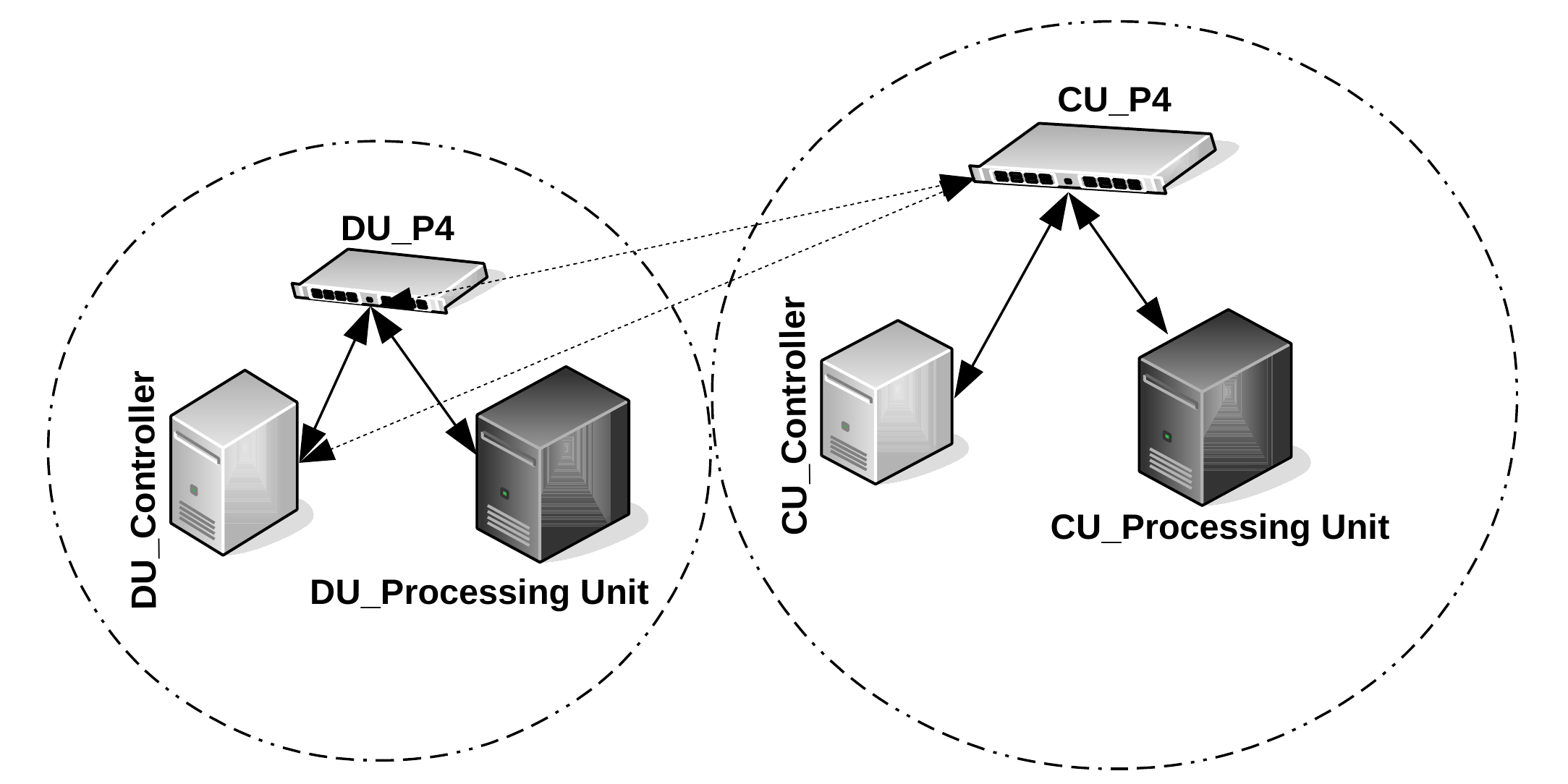}
    \caption{Proposed SMARTHO Framework.}\label{smarthofrm}
\end{figure} 

The main components of the proposed CU and DU architecture are COTS
compute servers, P4 switches, and a Network Controller. The compute
servers will implement the functions of CU and DU, P4 switches, and
Network Controller. The interconnections and components of SMARTHO
framework are shown in Figure~\ref{smarthofrm}.

The network controller at the CU (CU\_Controller) will store the ``UE
Mobility Information'' and the ``UE Context Information''. The network
controller at the DU (DU\_Controller) will store the RRC Connection
Reconfiguration (RRCCR) message. The P4 switches will process the
messages from processing units and perform the SMARTHO process, by
sending appropriate instruction messages to CU and DU Controllers.

The first handover of a given UE will set the UE context information
in the CU\_Controller. After the first HO is completed, the SMARTHO
initiation will happen which automates the subsequent handovers. The
P4 switches in CU (CU\_P4) and DU (DU\_P4) will send the instruction
messages to CU\_Controller to access the mobility information and
DU\_P4 switches to store the RRCCR message respectively.

\begin{figure}[hbtp]
  \centering
  \includegraphics[width=0.6\linewidth]{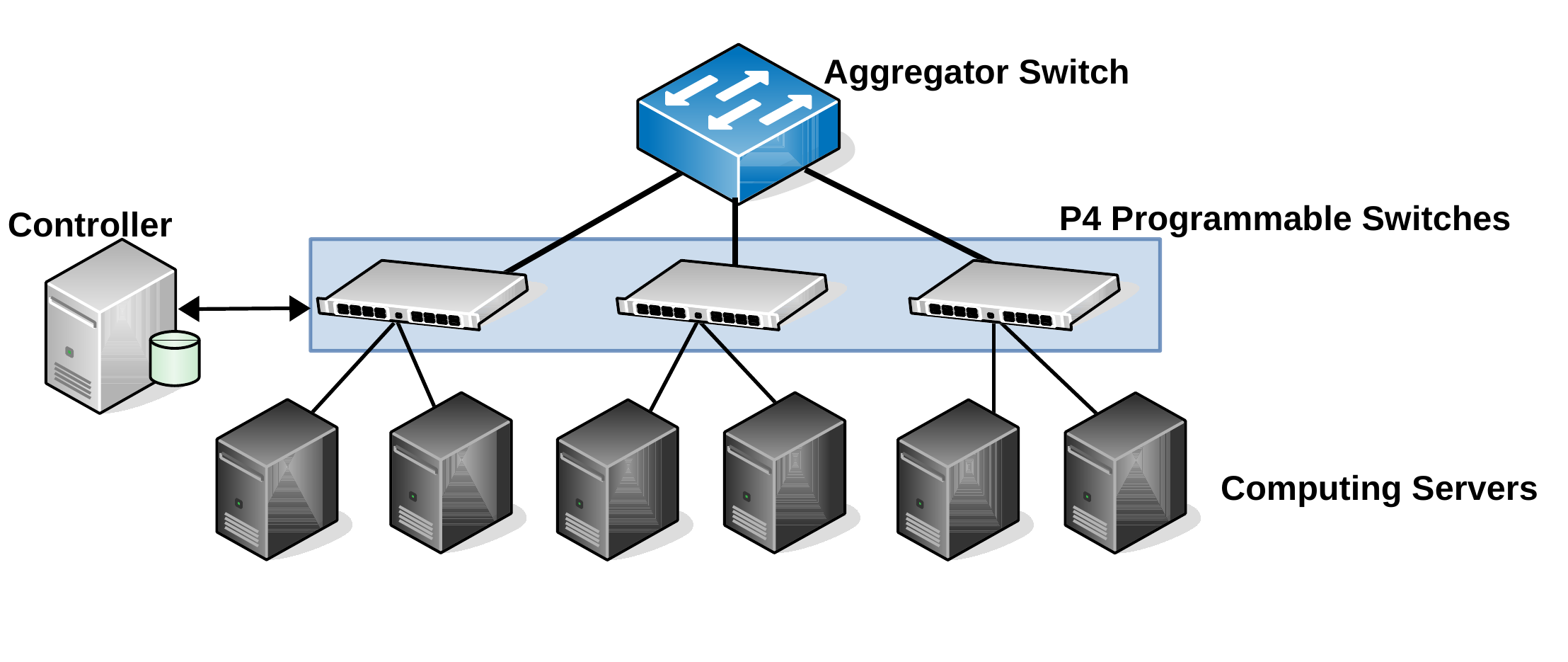}
    \caption{Topology interconnecting CU and
      DUs.}\label{CUDUArchitecture} 
\end{figure}

These P4 switches can be hardware switches\cite{benavcek2017line} or a
virtual switch\cite{P4Software}. Placement of P4 switches in CU and DU
can impact the routing performance of the system. A study of this
aspect is not in the scope of this report. Hence, without loss of
generality, we assume all the P4 switches are at the access layer
connected directly to servers and controller
as shown in  Figure~\ref{CUDUArchitecture}.

\subsection{Modified Handover Sequence}

\begin{figure}[htbp]
\centering
  \includegraphics[width=0.7\linewidth]{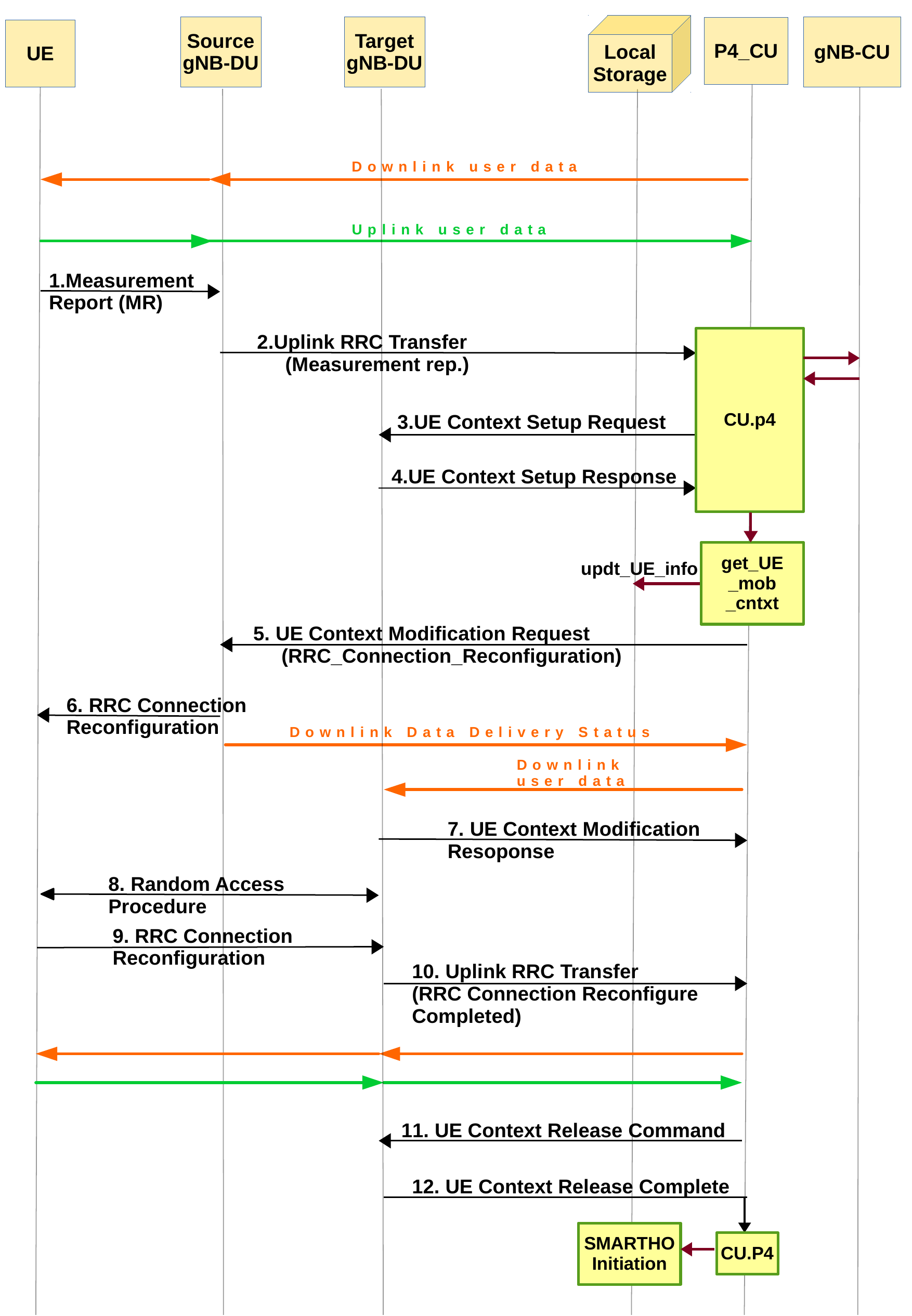}
    \caption{Sequence diagram of Intra-CU
      Handover.}\label{5gseqdiaintracuho}
\end{figure}

The entire 3GPP process with P4 switches in CU with sequence of
messages is shown in Figure~\ref{5gseqdiaintracuho}.
In the first handover, P4 switches will parse the incoming packets and
negotiate with local storage at CU to determine if the UE is having a
fixed path. If so, after the completion of first HO, the P4-switch
will generate "UE Context Setup Request" message and forward it to the
Target HO entities, on behalf of the UE.  This is referred to a Smart
Handover (SMARTHO) in this report. This action will trigger the HO
preparation phase, even before UE reaches the specified HO points, as
shown in Figure~\ref{5gintracuho1}.

This would make all the Target HO entities to reserve resources and
respond to CU with appropriate ``UE Context Setup Response''. The ``UE
Context Setup Response'' message would be saved at Source\_DU and can
be later forwarded by the P4-switch as a response to the UE MR. By
this spoofing approach, we parallelize the HO preparation phase, which
will improve the performance of the handover process.

\subsection{Architecture and Design of P4 switches} \label{designofp4}

There are several switch architectures such as
Pisces\cite{shahbaz2016pisces} and Portable Switch Architecture
(PSA)\cite{p416psa} that support protocol independent switches. In
this report, we use the Very Simple Switch (VSS)
Architecture\cite{P416}. VSS has basic programming blocks needed for
protocol independent switch, which are sufficient to implement the
SMARTHO process.

The programming blocks of VSS are: (i) Parser; (ii) Match-Action
Pipeline; and (iii) De-parser.  The parser is a Finite State Machine
(FSM), which either accepts or rejects the packet. For every packet
the P4 switch receives, it will parse the packets and would extract
the header information. The header information obtained is used in the
Match-Action Pipeline to invoke a necessary action routine in
Match-Action control block. The De-parser will reconstruct the packet,
putting back the extracted content of the header with necessary
modifications, if needed.
 
Next generation mobile networks have a complex packet structure.
Designing a parser for entire packet structure would overload the
functionality of the P4 switch, increasing the complexity of the
parser. Also, the structure of the packets for mobile networks would
depend on the state information. P4 switches are not scalable to parse
such packets as of now. To simplify this process, we design a
\textit{tag}-based approach to identify necessary packets for
SMARTHO. The tag will be added by the processing units or controller.
The P4-switches in the SMARTHO model handles three types of packets:

\begin{enumerate}
\item User packets of the 5G system: These packets are ICMP packets
  encapsulated over the tag, the forwarding is done using tag
  information. 

\item Control packets for HO: In case of Intra CU HO, the entire HO
  process has twelve control messages exchanging, shown in
  Figure~\ref{5gseqdiaintracuho}. These packets have to be identified
  and will be sent to P4 switches or controller for processing.

\item Instruction packets: These packets will either instruct the P4
  switch to initiate specific methods in Match-Action control block or
  the controller to store/retrieve the data.

\end{enumerate}

\subsection{Custom Data Structures}

Three special data structures have been defined to store the necessary
state information: Mobility Table (MT), Controller Cache (CC) and RRC
Table (RRCT). MT and CC will reside in CU\_Controller and RRCT will
reside in DU\_Controller. The details are given below.

\label{DSforarc}
A data structure is defined to store the necessary information needed for the SMARTHO process. We define three data structures Mobility Table (MT), Controller Cache (CC) and  RRC Table (RRCT). MT and CC will reside in CU\_Controller and RRCT will reside in DU\_Controller
\subsubsection{Mobility Table (MT)}
MT stores the mobility information of the UE. With the details in MT, P4 switch will identify the Target\_DU for the next HO. The controller would use MT information to trigger the SMARTHO-Initiation (discussed in the Section \ref{SMARTHOInit}) at an appropriate time. Every MT entry contains:
\begin{itemize}
\item UE-ID: Identification of the user equipment
\item Source DU ID: The source DU global identification
\item Target DU ID: The next target DU global identification for the current Source DU ID
\item Time Interval: Appropriate time interval after which the SMARTHO process is triggered.
\end{itemize}
\subsubsection{Controller Cache (CC)}
The UE Context Information is retrieved from the message "UE Context Setup Request", which is triggered from CU processing unit. This information thus retrieved is stored in CU Controller Cache (CC). CU\_P4 switch forwards the "UE Context Setup Request" to CU Controller as shown in Figure~\ref{SMARTHOStep1} to update the UE context information in CC. Every CC entry contains:
\begin{itemize}
\item UE-ID: Identification of the user equipment
\item UE-AMBR: Aggregated Max Bit Rate
\item UE-Security-Algorithm: Encryption algorithm used by UE
\item Security-Base-Key: Base key to encryption keys
\end{itemize}
\subsubsection{RRC Table (RRCT)}
RRCT will store the final HO preparation message (UEModReq/RRCCR) at DU\_Controller. The DU\_P4 switch will instruct the DU\_Controller to store the UEModReq message. The RRCT contains all the fields of UEModReq message, as shown below.
\begin{itemize}
\item UE-ID: Identification of the user equipment
\item Target DU ID: Aggregated Max Bit Rate
\item Bearer information: Bearer ID allocated by the Target\_DU
\item Security-Algorithm: Security algorithm at the Target DU
\end{itemize}

All the three types of packets are encoded with the respective
tags. The differentiation is done based on the extracted tag and
examining the valid/invalid bit\cite{P416}. The parser in P4 switch
should be indicated about the appropriate tag, for this we use,
Ethernet-Type from Ethernet header. IEEE802.3 has assigned EtherType
0x0101-0x01FF as experimental, we can use any of these for indication
of tag header. The parser routine of the P4 switch in CU is shown in
Algorithm~\ref{parsercup4}.

\begin{algorithm}
  \caption{parser block for cu.p4}\label{parsercup4}
  \begin{algorithmic}[1]
  \Statex \textit{header\_union } Tag\{ 
  \Statex \hspace{1cm}\textit{FrwdTag} t1;
  \Statex \hspace{1cm}\textit{CntrlTag} t2;
  \Statex \hspace{1cm}\textit{InstTag} t3;\}
  \Statex \textit{struct } Parsed\_packet \{
  \Statex \hspace{1cm}\text{Ethernet }ethernet;
  \Statex \hspace{1cm}\text{Tag } tag;\}
  \Statex \textbf{parser}\text{ Simple\_Parser}\text{($packet\_in$ packet,}
  \Statex \hspace{3.2cm} \text{$out$ $Parsed\_packet$ hdr)}\{     
  \Statex \hspace{0.5cm}$state$ start 
  \Statex \hspace{1cm}packet.$extract$(hdr.ethernet);
  \Statex \hspace{1cm}$transition$ $select$(hdr.ethernet.etherType)
  \Statex \hspace{1.5cm}$16w0$x$0101$ : parse\_inst\_tag;
  \Statex \hspace{1.5cm}$16w0$x$0102$ : parse\_cntrl\_tag;
  \Statex \hspace{1.5cm}\text{$default    $} : parse\_frwd\_tag;
  \Statex \hspace{0.5cm}$state$ parse\_inst\_tag 
  \Statex \hspace{1cm}packet.$extract$(hdr.tag.t3);
  \Statex \hspace{1cm}$transition$ $accept;$
  \Statex \hspace{0.5cm}$state$ parse\_cntrl\_tag 
  \Statex \hspace{1cm}packet.$extract$(hdr.tag.t2);
  \Statex \hspace{1cm}$transition$ $accept;$
  \Statex \hspace{0.5cm}$state$ parse\_frwd\_tag 
  \Statex \hspace{1cm}packet.$extract$(hdr.tag.t1);
  \Statex \hspace{1cm}$transition$ $accept;$
  \Statex \hspace{1cm}\}
  \end{algorithmic}
\end{algorithm}

\section{Implementation Details} \label{smarthoimpl}

The HO preparation is a resource allocation phase, in the case of
fixed path mobile devices the resource allocation can be done a
priori. The idea is to preset all the subsequent HOs with appropriate
timing delays based on the first HO request. 

The preparation phase for the second Intra-CU HO is done before the UE
reaches the second Intra-CU HO point. The P4 switch initiates the
preparation phase for the Second Intra CU HO, i.e., CU\_P4 switch
along with CU\_Controller spoofs the UE and sends a ``UE Message Setup
Request'' to the Target\_DU. When the UE reaches the vicinity of the
second HO point, UE will trigger the MR message to Source\_DU;
subsequently, the Source\_DU\_P4 will respond with the RRCCR message.

\begin{figure}[hbtp]
  \centering
  \includegraphics[width=0.7\linewidth]{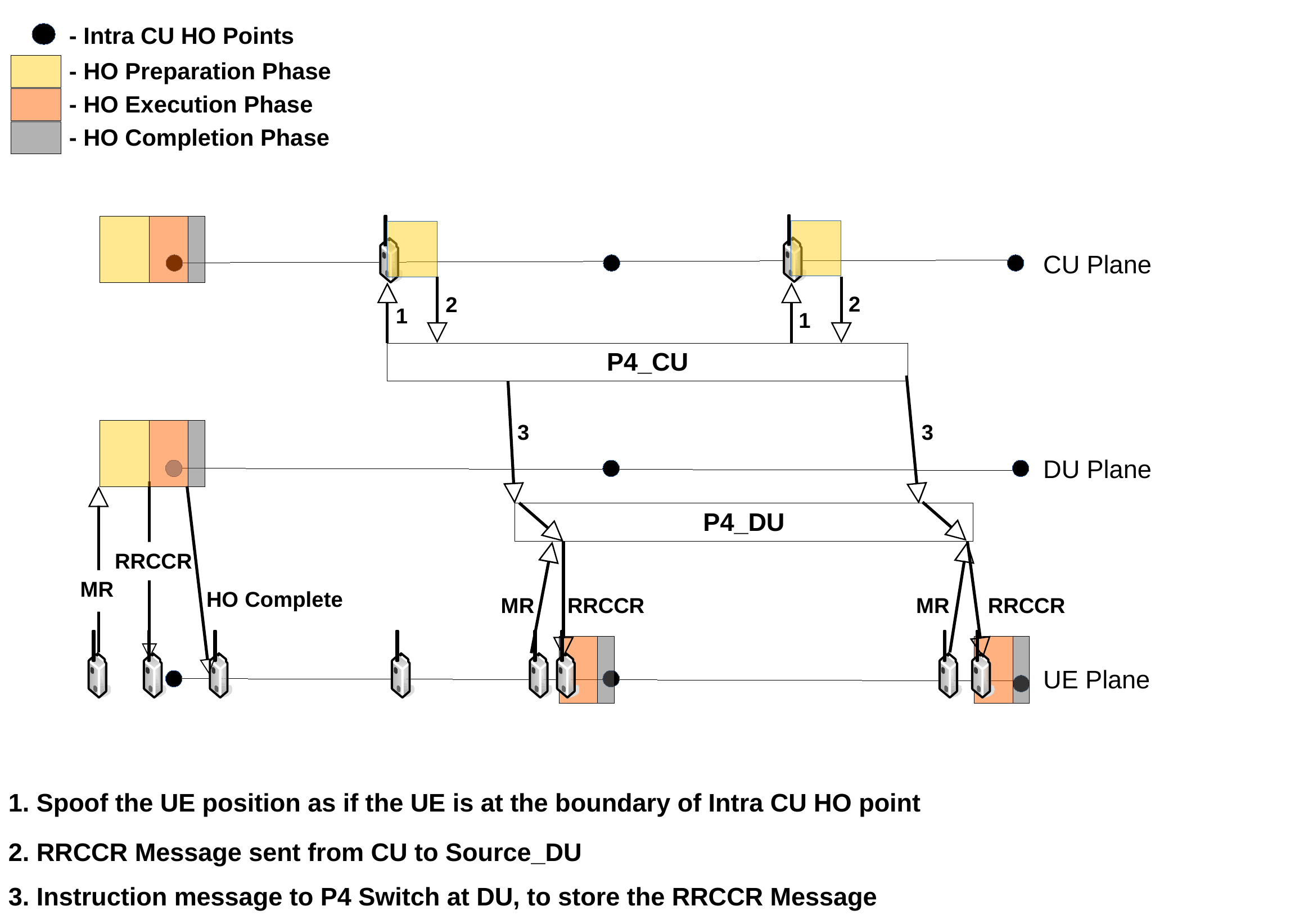}
    \caption{Operation of handover process, using P4 switches.}\label{5gintracuho2}
\end{figure}

As described earlier, we perform the HO preparation phase in advance
of the UE movement, in order to decrease the overall HO time.
Figure~\ref{5gintracuho2} presents the working details of SMARTHO,
with a sequence of three Intra-CU handover (HO) points. 
The operation of SMARTHO has three phases: SMARTHO-Data Setup,
SMARTHO-Initiation and SMARTHO-Completion, as described below.

\begin{figure}[hbtp]
  \centering
  \includegraphics[width=0.6\linewidth]{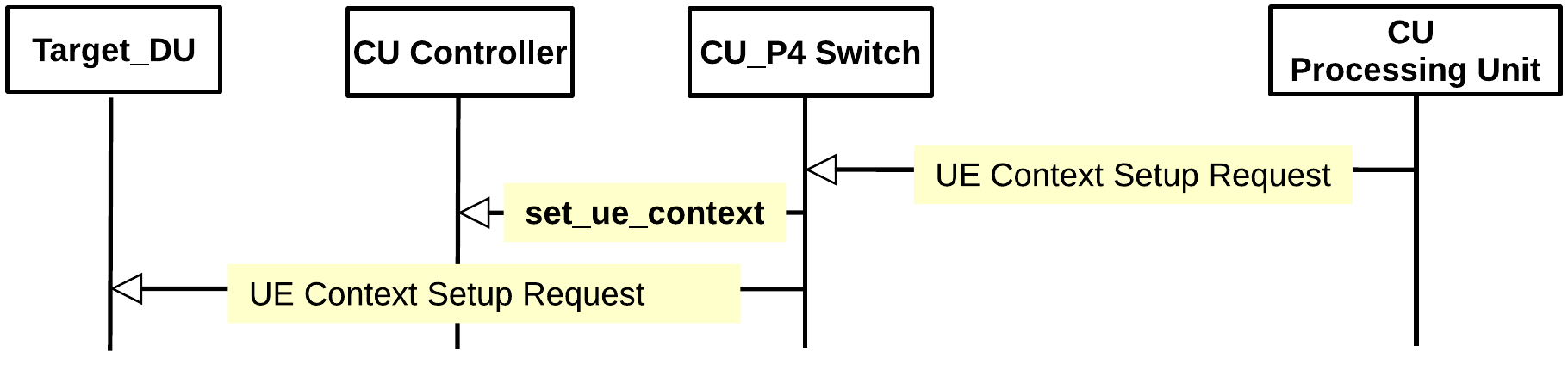}
    \caption{Trigger sequence of SMARTHO.}\label{SMARTHOStep1}
\end{figure} 

\subsection{Data Setup}

The current context of the UE has to be retrieved, before the start of
the SMARTHO process. The context information of UE can be retrieved
from the ``UE Context Setup Request'' message, which is exchanged
between CU and Target\_DU as shown in
Figure~\ref{5gseqdiaintracuho}. The UE context information is updated
in the data table CC.

This message is sent to the CU\_P4 switch. The CU\_P4 switch can
identify the control packets for HO, this can be done by changing the
code at CU part, to send the HO message ``UE Context Setup Request''
with tag value $0$x$03$, as discussed in Section \ref{designofp4}. The
CU\_P4 will identify the tag and execute a routine to send the message
set\_ue\_context to the CU\_Controller, which will store the UE
context information in CC, as shown in Figure~\ref{SMARTHOStep1}.
The set\_ue\_context contains the UE identifier, Aggregate Maximum Bit
Rate (AMBR) for the UE, and other relevant information.

\begin{algorithm}
 \caption{cu.p4}\label{cup4}
 \begin{algorithmic}[1]
 \Statex \textbf{control}\text{ Ingress}\text{($inout$ $headers$ hdr,}
 \Statex \hspace{1cm} $inout$ $metadata$ meta,
 \Statex \hspace{1cm} \text{$inout$ $standard\_metadata\_t$ standard\_metadata) \{}
 \Statex \hspace{0.5cm} $table$ etherforward 
 \Statex \hspace{1cm} $key$ = hdr.ether.dst\_addr : $exact$;
 \Statex \hspace{1cm} $actions$ = 
 \Statex \hspace{1.5cm} ether\_port\_forward;
 \Statex \hspace{1.5cm} operation\_drop;
 \Statex \hspace{1.5cm} $const$ default\_action = operation\_drop();
 \Statex \hspace{0.5cm} action cu\_controller\_forward()
 \Statex \hspace{1cm} $standard\_metadata.egress\_spec$ 
 \Statex \hspace{1.5cm}  $= controller\_port$;
 \Statex \hspace{0.5cm} $table$ source\_gnb\_controller\-forward 
 \Statex \hspace{1cm} $key$ = hdr.ue\_context.src\_gnb\_addr : $exact$;
 \Statex \hspace{1cm} $actions$ = 
 \Statex \hspace{1.5cm} prepare\_\_port\_forward;
 \Statex \hspace{1.5cm} operation\_drop;
 \Statex \hspace{1.5cm} $const$ default\_action = operation\_drop();
 \Statex \hspace{0.5cm} \textit{$apply$}\{   
 \Statex \hspace{1cm}  if (hdr.tag.$isValid()$) 
 \Statex \hspace{1.5cm}  if(hdr.ue\_context.$isValid()$)
 \Statex \hspace{2cm}    cu\_controller\_forward();
 \Statex \hspace{1.5cm}  else 
 \Statex \hspace{2cm}    source\_gnb\_controller\_forward.$apply()$;
 \Statex \hspace{1cm}         else 
 \Statex \hspace{1.5cm}            etherforward.$apply()$;
 \Statex \hspace{1.4cm} \}
 \Statex \}  
 \end{algorithmic}
\end{algorithm}

The P4 switch at CU identifies the set\_ue\_context message and
forwards it to the CU\_Controller, this is shown at a high level in
Algorithm~\ref{cup4}. Once the CU\_Controller receives the
set\_ue\_context message, it updates its CC using a packet sniffer at
the controller.


\begin{figure}[hbtp]
  \centering
  \includegraphics[width=0.6\linewidth]{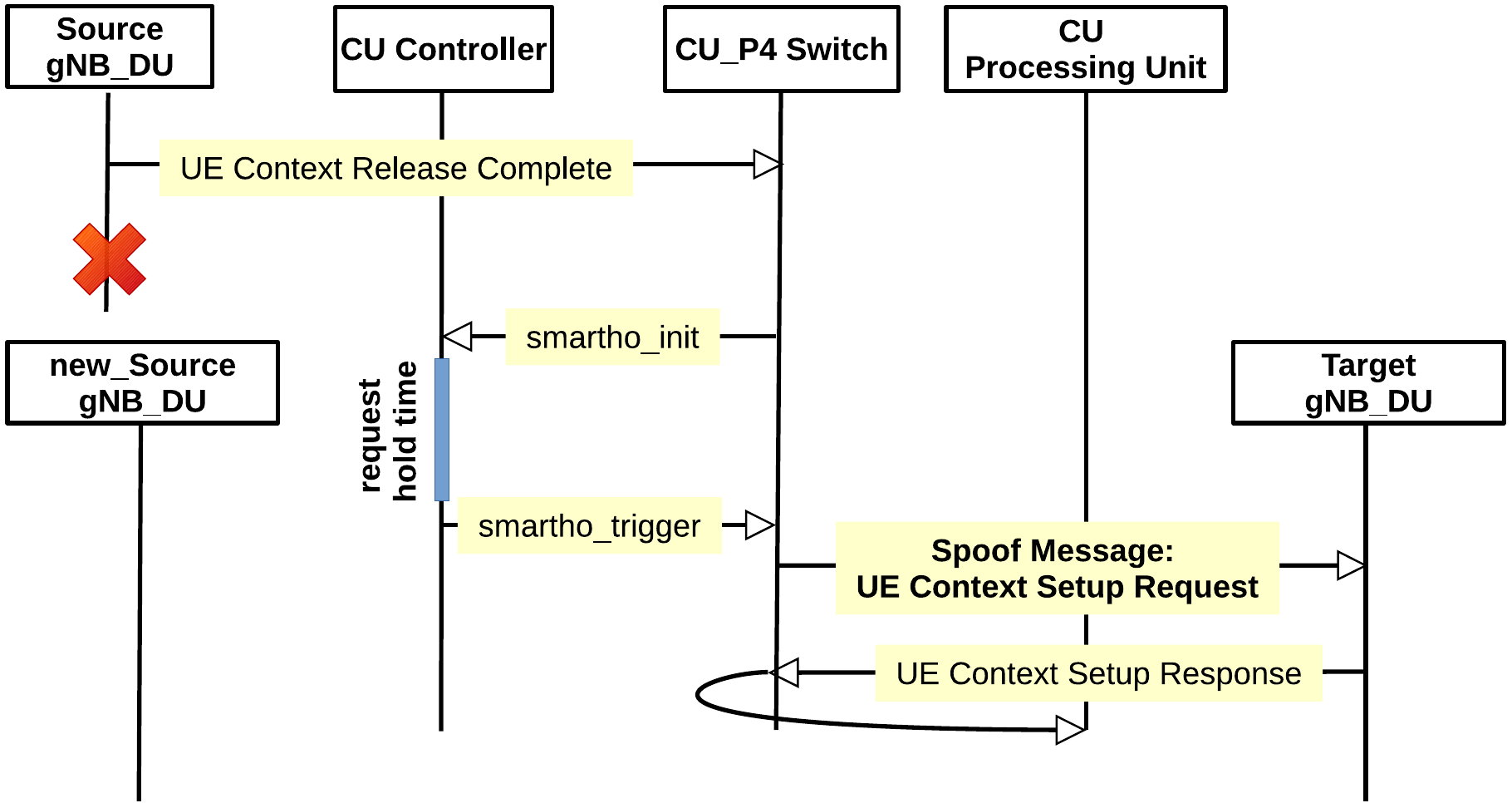}
    \caption{Initiation sequence of SMARTHO.}\label{SMARTHOStep2}
\end{figure} 

\subsection{SMARTHO - Initiation} \label{SMARTHOInit}

\algdef{SE}[VARIABLES]{Variables}{EndVariables}
   {\algorithmicvariables}
   {\algorithmicend\ \algorithmicvariables}
\algnewcommand{\algorithmicvariables}{\textbf{global variables}}
%
%

The initiation of the SMARTHO process is shown in
Figure~\ref{SMARTHOStep2}. The Source\_gNB\_DU sends the ``UE Context
Release Complete'' message with a tag value of $0$x$0c$ to the
CU\_P4. This switch parses the packet and identifies the message with
the tag value and initiates the process of SMARTHO. This is done by
sending the smartho\_init message to the CU\_Controller with a tag
value of $0$x$02$. The purpose of the smartho\_init message is to
retrieve the address of Target\_gNB\_DU from MT for the next HO and
\textit{delay} information of the UE. This delay value is used to hold
the process before starting the preparation phase.

The CU\_Controller runs a packet sniffer at the ingress port. When a
smartho\_init message is received, the sniffer runs a background
process. This will send the smartho\_trigger message to the CU\_P4
switch with a tag value of $0$x$02$ as shown in
Algorithm~\ref{smarthoInit}. The smartho\_trigger message is sent
after a particular delay value, as discussed later in
Section~\ref{timinganalysis}.

The smartho\_trigger message is the basis to send the spoofed
``UE Context Setup Request'' message for the next HO to the
Target\_gNB\_DU. This will initiate the HO preparation phase for the
subsequent HO.

\begin{algorithm}
\caption{smarthoInit}\label{smarthoInit}
\begin{algorithmic}[1]
\Variables
  \State $mobility\_tag=2$
\EndVariables
\Function{TriggerSmartho}{ue\_id,src\_du}
\State mobility\_details[]=query\_mobility\_table(ue\_id,src\_du)
\State context\_details[]=query\_controller\_cache(ue\_id)
\State $delay$(mobility\_details[time\_interval])
\State ether=Ether(dst\_addr, type=0x0101)
\State tag=Tag(mobility\_tag)
\State context\_info=create\_header(context\_details)
\State ue\_context\_req\_pkt =  ether/tag/context\_info
\State srp1(ue\_context\_req\_pkt, iface="eth")
\EndFunction
\end{algorithmic}
\end{algorithm}

\begin{figure}[hbtp]
  \centering
  \includegraphics[width=0.6\linewidth]{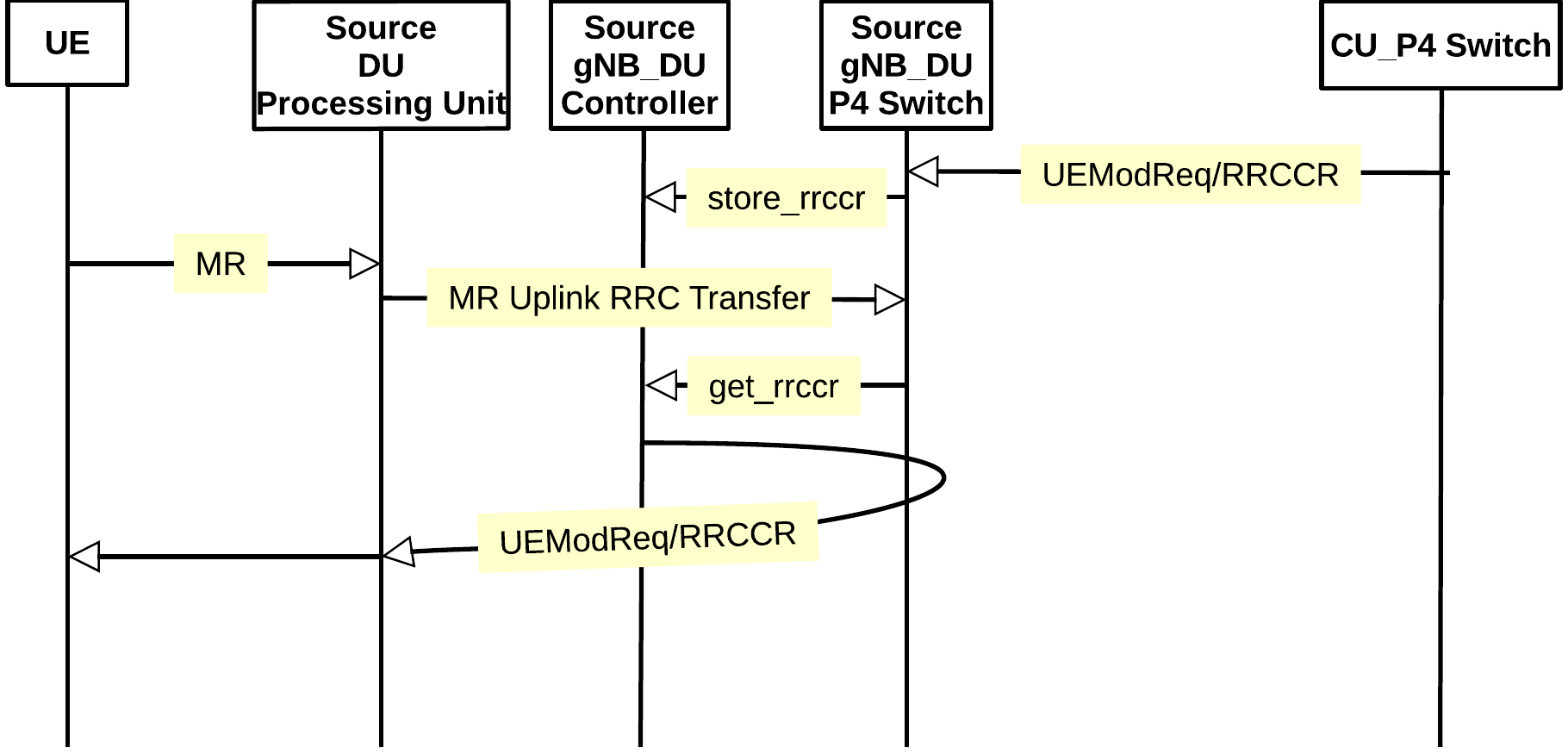}
    \caption{Completion sequence of SMARTHO.}\label{SMARTHOStep3}
\end{figure} 

\subsection{SMARTHO Completion}

The final phase of SMARTHO is to handover the UEModReq/RRCCR message
as a response to UE MR, as shown in Figure~\ref{SMARTHOStep3}.  The
UEModReq/RRCCR message that is sent from CU to Source\_DU is
intercepted by the Source\_DU\_P4 switch. This would instruct the
Source\_DU\_Controller to store UEModReq/RRCCR message. This message
contains the UEModReq information that is updated in the RRCT of
DU\_Controller. Algorithm~\ref{ducontroller} and the P4 code segment
shown in Algorithm~\ref{dup4} present the details of this operation.

When a UE sends the MR to Source\_DU, the Source\_DU would respond
with ``Uplink RRC Transfer message'' to CU. The DU\_P4 switch 
intercepts this message and instructs the controller to get the
UEModReq/RRCCR message which is forwarded to UE as shown in
Figure~\ref{SMARTHOStep3}. 

\begin{algorithm}
\caption{DUController}\label{ducontroller}
\begin{algorithmic}[1]
\Variables
 \State $store\_rrc\_tag=15$
 \State $mr\_uplink\_rrc\_tag=1$
\EndVariables
\Function{Data\_Updt}{packet}
\If {$packet.tag$ == $store\_rrc\_tag$} 
\State rrc\_packet\_data[]=extract\_packet\_content(packet)
\State query load rrct rrc\_packet\_data[]
\EndIf
\If {$packet.tag$ == $mr\_uplink\_rrc\_tag$} 
\State query uemod\_reqmsg=get rrct(packet.ue)
\State srp1(uemod\_reqmsg, iface="eth")
\EndIf
\EndFunction
\Procedure{main}{}
\State $\text{sniff(iface="eth",prn=DATA\_UPDT)}$
\EndProcedure
\end{algorithmic}
\end{algorithm}

\begin{algorithm}[H]
\caption{du.p4}\label{dup4}
\begin{algorithmic}[1]
\Variables
\Statex $ue\_context\_tag=1$
 \Statex $mobility\_tag=2$
\EndVariables
\Function{Parser}{$packet\_in$ packet,$out$ $headers$ hdr}
 \Statex \textbf{parser}\text{ Simple\_Parser}\text{($packet\_in$ packet,}
 \Statex \hspace{3.2cm} \text{$out$ $headers$ hdr)}\{     
 \Statex \hspace{0.5cm}$state$ start \{
 \Statex \hspace{1cm}packet.$extract$(hdr.ethernet);
 \Statex \hspace{1cm}$transition$ $select$(hdr.ethernet.etherType) \{
 \Statex \hspace{1.5cm}$0x0101$ : parse\_inst;
 \Statex \hspace{1.5cm}$default$      : $accept$; \hspace{3cm}\}
 \Statex \hspace{2cm} \}
 \Statex \hspace{0.5cm}$state$ parse\_inst \{
 \Statex \hspace{1cm}packet.$extract$(hdr.tag);
 \Statex \hspace{1cm}$transition$ $select$(hdr.tag.tag\_value) \{
 \Statex \hspace{1.5cm}$0x01$ : parse\_ue\_context;
 \Statex \hspace{1.5cm}$default$      : $accept$; \hspace{3cm}\}
 \Statex \hspace{2.7cm} \}
 \Statex \hspace{0.5cm}$state$ parse\_ue\_context \{
 \Statex \hspace{1cm}packet.$extract$(hdr.ue\_context);
 \Statex \hspace{1cm}$transition accept;$
 \Statex \hspace{2.7cm} \}
 \Statex \}
\EndFunction
\end{algorithmic}
\end{algorithm}

\subsection{Delay Estimation for Early Resource
  Allocation} \label{timinganalysis} 

The UE context setup is done by the Target\_DU before allocating the
resources, as described earlier. Once the UE context set-up is done at
the T\_DU, the T\_DU waits for the ``Random Access Procedure''.  If
this is not received before timer expiry, the ``UE Context Release
Request'' will be initiated to release all the necessary bearers. The
timer expiry is triggered based on the user inactivity or by policy
controls\cite{ralfkreherltesignaling}. In SMARTHO, the advanced
allocation of resources would be wasted. Hence, an appropriate delay
has to be put before SMARTHO Initiation.

To estimate the delay ($t_{delay}$) to initiate the SMARTHO process,
we need three inputs: 
(i) Estimated arrival of Measurement Report (MR) for next HO ($t_{MR}$);
(ii) Total Response time for HO preparation ($t_{prep\_HO}$);
(iii) Trigger time, for "UE Context Release Request" by T\_DU ($t_{trig}$)

Using Machine Learning techniques with the features such as traffic
intensity at switches, history information and so on, we can predict
the estimated arrival time of the MR message. 

The HO preparation time ($t_{prep\_HO}$) would include the processing
times of CU and DU cloud units and processing times of routers
connecting CU, DU and RRH. For estimating this we model the system as
a simple network of queues. We assume that the packet arrival process
at a UE is Poisson; service time is exponential; and routers have
limited buffer capacity. We model the routers as a $M/M/1/B$ queue,
and the CU and DU entities as $M/M/1$. We model the system as a tandem
of $M/M/1/B$ and $M/M/1$ queuing system. The variables are shown in
Table~\ref{Variables_queuing_model}.

\begin{table}[htbp]
\caption{Variables in queuing model}
\label{Variables_queuing_model}
\begin{center}
 \begin{tabular}{|l|l|} 
 \hline
 $t_{pd\_sDU\_CU}$ & Propagation delay from Source\_DU to CU  \\ 
 \hline
 $t_{pd\_tDU\_CU}$ & Propagation delay from Target\_DU to CU  \\ 
 \hline
 $t_{pc\_cd}$ & \begin{tabular}[c]{@{}l@{}} Expected response time at CU and DU\\ in HO preparation phase \end{tabular}\\
 \hline
 $t_{pc\_rt}$ & \begin{tabular}[c]{@{}l@{}}Expected delay by routers in HO preparation phase\end{tabular}  \\
 \hline
 $n^{r\_sd}$ &\begin{tabular}[c]{@{}l@{}}number of routers between RRH and Source\_DU \end{tabular}  \\
 \hline
 $n^{r\_td}$ &\begin{tabular}[c]{@{}l@{}}number of routers between RRH and Target\_DU \end{tabular} \\
 \hline
 $n^{sd\_cu}$ &\begin{tabular}[c]{@{}l@{}}number of routers between Source\_DU and CU \end{tabular} \\
  \hline
 $n^{td\_cu}$ &\begin{tabular}[c]{@{}l@{}}number of routers between Target\_DU and CU \end{tabular} \\
   \hline
 $n$ & \begin{tabular}[c]{@{}l@{}} total number of routers between RRH, Source\_DU,\\ Target\_DU and CU. Each router indexed as $x \epsilon \{1...n\}$\\ $ n = n^{td\_cu}+n^{sd\_cu}+n^{r\_sd}+n^{r\_sd}$\end{tabular}  \\ 
  \hline
 $B_x$ & \begin{tabular}[c]{@{}l@{}}Buffer size in router $x$, present between \\ CU and DU, $ x \epsilon {1,2,...n}$\end{tabular}  \\ 
  \hline
 $\lambda_{x}$ & packet arrival rates in router $x$  \\
  \hline
 $\mu_{x}$ & router $x$ processing rates in  \\
  \hline
 $E[r_{x}]$ & expected response time of router $x$  \\
  \hline
 \end{tabular}
\end{center}
\end{table}

\begin{figure*}[htbp]
\begin{subfigure}[b]{0.33\textwidth}
        \includegraphics[width=0.9\textwidth]{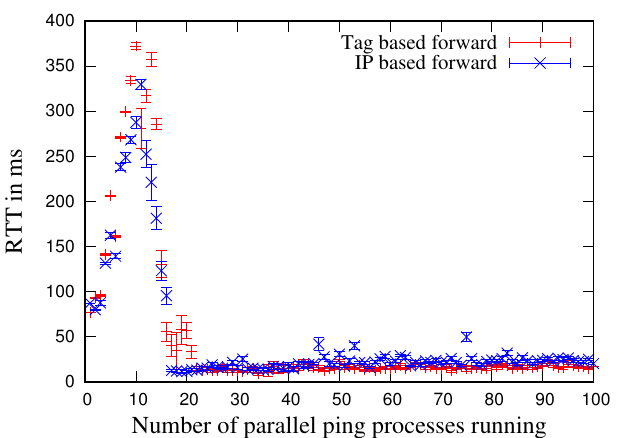}
        \caption{Hosts connected with 1-transit switch}
        \label{tagtimeperf_1}
\end{subfigure}
\begin{subfigure}[b]{0.33\textwidth}
        \includegraphics[width=0.9\textwidth]{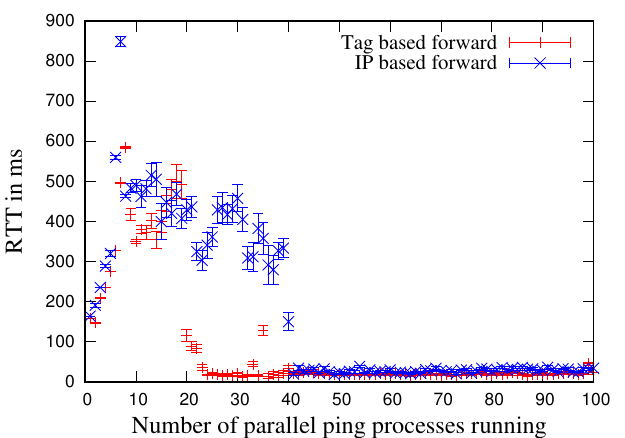}
        \caption{Hosts connected with 2-transit switches}
        \label{tagtimeperf_2}
\end{subfigure}
\begin{subfigure}[b]{0.33\textwidth}
        \includegraphics[width=0.9\textwidth]{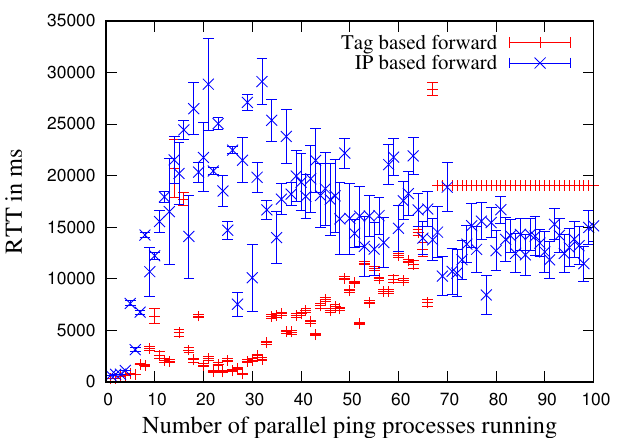}
        \caption{Hosts connected with 4-transit switches}
        \label{tagtimeperf_4}
\end{subfigure}
\begin{subfigure}[b]{0.33\textwidth}
        \includegraphics[width=0.9\textwidth]{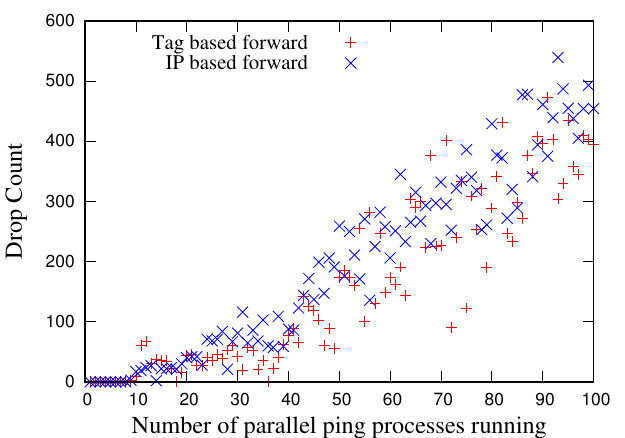}
        \caption{Hosts connected with 1-transit switch}
        \label{tagdropperf_1}
\end{subfigure}
\begin{subfigure}[b]{0.33\textwidth}
        \includegraphics[width=0.9\textwidth]{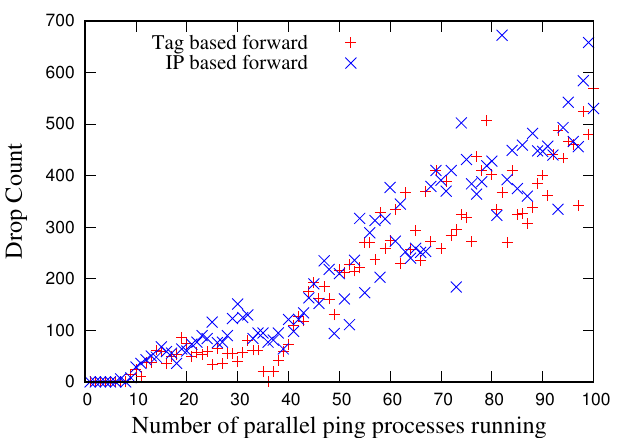}
        \caption{Hosts connected with 2-transit switch}
        \label{ftagdropperf_2}
\end{subfigure}
\begin{subfigure}[b]{0.33\textwidth}
        \includegraphics[width=0.9\textwidth]{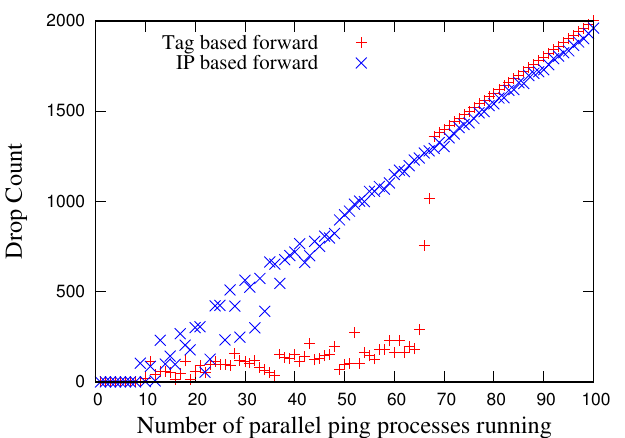}
        \caption{Hosts connected with 4-transit switch}
        \label{tagdropperf_4}
\end{subfigure}
\caption{Comparison of Tag- and IP-based forwarding mechanisms.} 
\label{tagperf}
\end{figure*}

For $M/M/1/B_x$ system, the response time is given by:

\begin{equation}\label{eq:1}
E[r_{x}] = \frac{\lambda_x}{\mu_x-\lambda_x} + \frac{B_x\lambda_x^{B_x+1}}{\mu_x(\mu_x^{B_x}-\lambda_x^{B_x})}  
\end{equation}

For CU and DU as $M/M/1$, the steady-state response time is given by:
\begin{equation}\label{eq:2}
E[r_{X}] = \frac{1}{\mu_{X}-\lambda_{X}} \text{ where, $X\varepsilon$\{CU,S\_DU,T\_DU\}}
\end{equation}

The total time taken for HO preparation is processing the Context
Requests and Measurement report. The four messages indexed 2,3,4,5
shown in Figure~\ref{5gseqdiaintracuho} are HO preparation
messages.\\ The processing time taken by the routers in HO preparation
phase ($t_{proc\_{rt}}$) is,
\begin{align*}
t_{proc\_{rt}} = 2\left(\sum_{x=1}^{n^{sd\_cu}}E[r_{x}] + \sum_{x=1}^{n^{td\_cu}}E[r_{x}]\right)
\end{align*}
The processing time taken by the CU and DU in HO preparation phase ($t_{proc\_{cd}}$) is,
\begin{align*}
t_{proc\_{cd}} = 2*\left(E[r_{S\_DU}] + E[r_{T\_DU}]\right)
\end{align*}
Total time taken for HO preparation is,
\begin{align*}
\small
t_{prep\_HO} = 2*t_{pd\_sDU\_CU} + 2*t_{pd\_tDU\_CU} + t_{pc\_{rt}} + t_{pc\_{cd}}
\end{align*}

The trigger time ($t_{trig}$) will include the trigger time and uplink
transfer time, approximated as:

\begin{align*}
t_{trig} = \text{trigger time}+t_{pd\_tDU\_CU}+\sum_{x=1}^{n^{td\_cu}}E[r_{x}])
\end{align*}
\begin{align*}
t_{delay} = t_{MR}-(t_{prep\_HO}-t_{trig})
\end{align*}

This value of delay of $t_{delay}$ is used an approximate value during
SMARTHO initiation, described earlier in Section~\ref{SMARTHOInit}.

%
%
%
%

\section{Implementation in Mininet Emulator}





The proposed SMARTHO framework was implemented in the mininet
emulation environment \cite{mininet}, where mininet-based hosts
emulate the CU and DU. Mininet hosts are connected using P4 switches,
developed using the P4 behaviour model (P4BM) with VSS model
architecture, \cite{P4Software}.  Raw data packets are created using
the \textit{scapy} tool\cite{scapy}, that sends a continuous sequence
of raw data packets from one host to another. User and control traffic
are also generated to simulate the mobile traffic and measure the HO
performance. User traffic is represented using ICMP ping packets over
a tag. The measurement of IP and tag based forwarding is done on user
traffic. Control traffic is generated to simulate the HO procedure,
packets are created with customized headers containing UE
identification, over the tag. The tag of the control packet is also
used as the identification for the HO message.

\begin{figure}[htbp]
  \centering
  \includegraphics[width=0.6\linewidth]{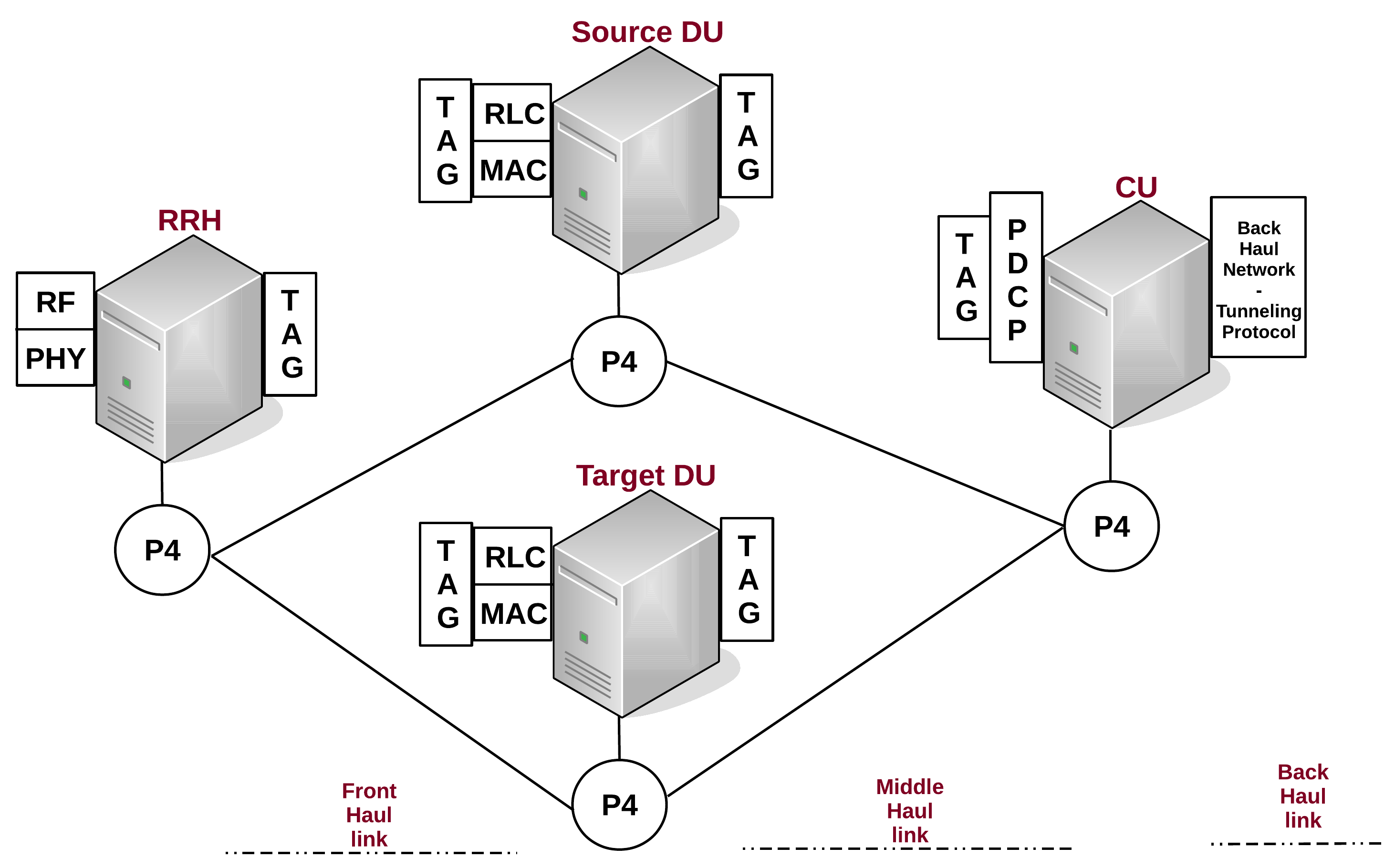}
    \caption{Network topology for Performance study of
      SMARTHO.}\label{simarch}  
\end{figure}

\subsection{Comparison of Tag and IP-based forwarding}

In order to study tag- and IP-based forwarding, user traffic is sent
among the hosts. P4 switches between these hosts parse the packets
and either forward the packet or execute the SMARTHO process. This
kind of tag-based approach is already investigated by Fayazbakhsh et
al. \cite{fayazbakhsh2014enforcing}, where they used the tag for
origin binding.

The comparison results are shown in Figure~\ref{tagperf}, with hosts
separated by one, two or four intermediate switches. The metrics
measured are the average response time and drop count of the
packets. As seen, tag-based forwarding performs better than IP based
forwarding.  Consider Figure~\ref{tagdropperf_4} and x-axis range of
(20,60) parallel ping process. Here, it is clearly seen that tag-based
forwarding is showing much lower packet drops when the number of hops
increase.  

In specialized environments such mobile networks, which are not
connected to the Internet until the Packet Gateway, a tag-based
forwarding approach is better. The tag-based identification of packets
makes the P4 parser simple, allowing innovations in other aspects too,
such as network slicing.

\begin{figure}[hbtp]
  \centering
  \includegraphics[width=0.6\linewidth]{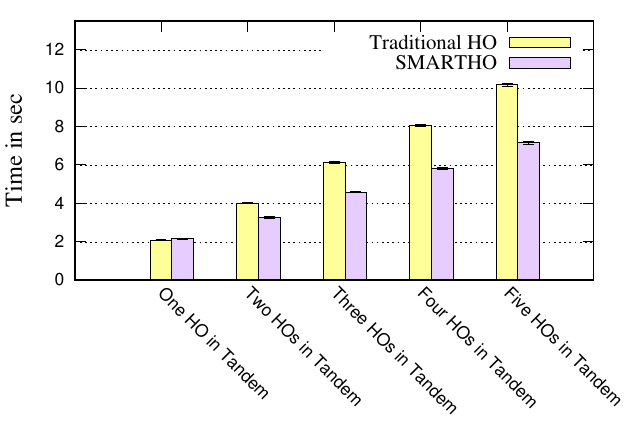}
    \caption{Performance of Intra-CU HOs in tandem.}\label{singHOperfintand}
\end{figure}

\subsection{Performance of SMARTHO handover}

\begin{figure*}[htbp]
\begin{subfigure}[b]{0.33\textwidth}
        \includegraphics[width=0.9\textwidth]{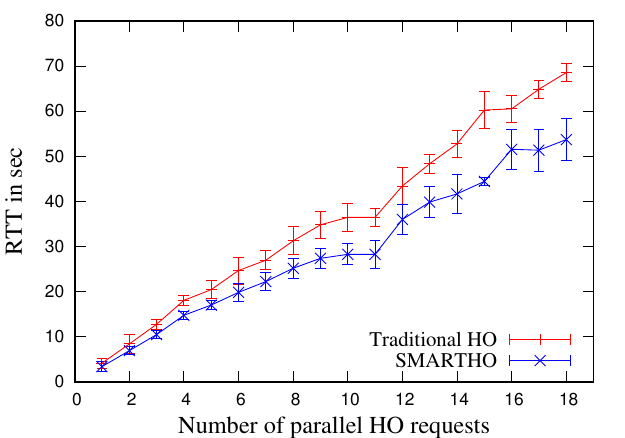}
        \caption{Two Intra-CU HO sequence}
        \label{SMARTHO_2_Resp}
\end{subfigure}
\begin{subfigure}[b]{0.33\textwidth}
        \includegraphics[width=0.9\textwidth]{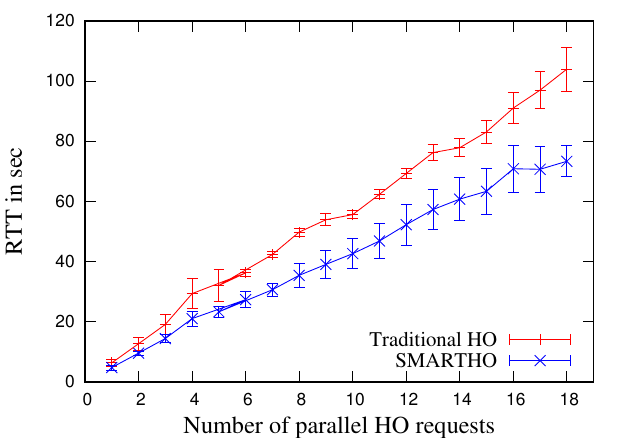}
        \caption{Three Intra-CU HO sequence}
        \label{SMARTHO_3_Resp}
\end{subfigure}
\begin{subfigure}[b]{0.33\textwidth}
        \includegraphics[width=0.9\textwidth]{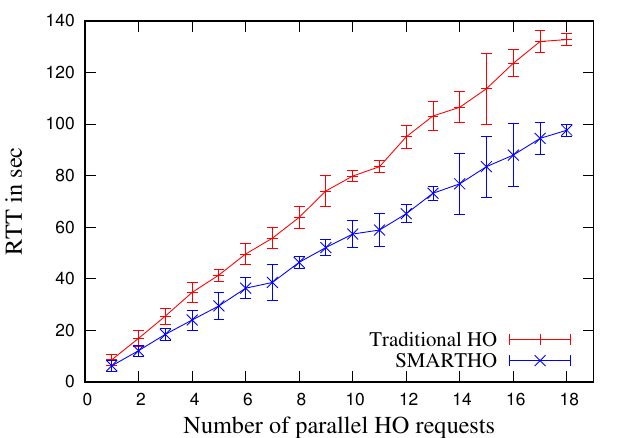}
        \caption{Four Intra-CU HO sequence}
        \label{SMARTHO_4_Resp}
\end{subfigure}
\caption{Performance comparison of traditional handover and SMARTHO in
  terms of Intra CU-HO time. }
\label{SMARTHO_Resp1}
\end{figure*}

\begin{figure*}[htbp]
\begin{subfigure}[b]{0.33\textwidth}
        \includegraphics[width=0.9\textwidth]{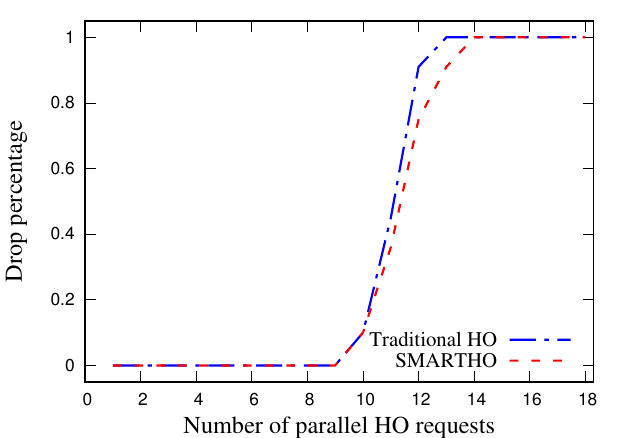}
        \caption{Two Intra-CU HO sequence}
        \label{drop_per_2}
\end{subfigure}
\begin{subfigure}[b]{0.33\textwidth}
        \includegraphics[width=0.9\textwidth]{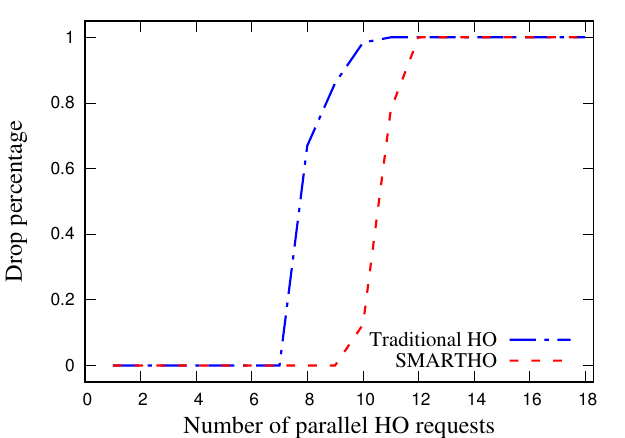}
        \caption{Three Intra-CU HO sequence}
        \label{drop_per_3}
\end{subfigure}
\begin{subfigure}[b]{0.33\textwidth}
        \includegraphics[width=0.9\textwidth]{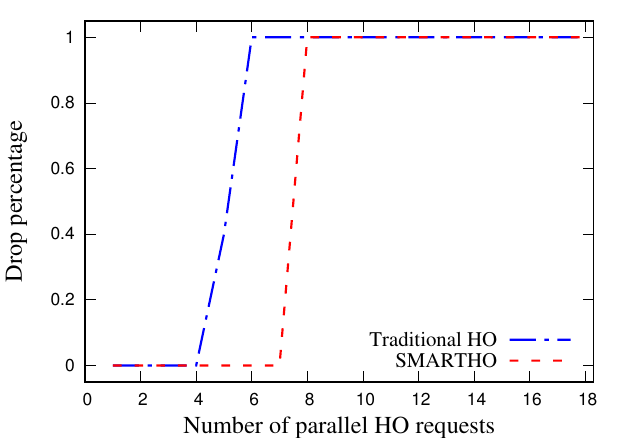}
        \caption{Four Intra-CU HO sequence}
        \label{drop_per_4}
\end{subfigure}
\caption{Analysis of handover failure percentages.}
\label{SMARTHO_Resp2}
\end{figure*}

For this study, a mininet environment as shown in Figure~\ref{simarch}
was created. We considered a tandem of Intra-CU handovers, sending
user packets between the RRH and CU. User packets are generated as
parallel ping process in RRH to simulate varying arrival rates. The
inter-arrival time between Intra-CU HO was exponential. The HO
procedure begins at the RRH by sending MR message to Source\_DU as
shown in the simulation architecture. The HO time is measured from the
moment RRH has sent the MR message to the Source\_DU, to the RRCCR
message received at RRH indicating the HO is completed.

Figure~\ref{singHOperfintand} presents the performance for handover
time on a single UE. The graph shows the total time spent for
handover. As seen, the SMARTHO process performs better than the
traditional HO process.  There is no improvement of HO response time
with single HO, this is because the SMARTHO process will perform the
data setup in first HO and automates the subsequent HOs. Improvement
of 18\% for two tandem HOs and 25\% for three tandem HOs is achieved
and this improvement will increase as the tandem of HOs
increases. This is because the overall time spent on HOs will
proportionally decrease as the HO preparation phase is done in advance
for all the subsequent HOs.

In the next study, we increased the intensity of HO requests, with
multiple UEs requesting handovers. Figure~\ref{SMARTHO_Resp1}
presents the response time. The results show that the SMARTHO process
performs better than the traditional HO process, with higher
improvements with increase in the number of transmit nodes.

Figure~\ref{SMARTHO_Resp2} presents the drop percentage of the HOs,
where a handover is considered dropped when the response exceeds a
threshold. It is observed that the proposed SMARTHO process is better
when the number of intermediate nodes is higher.

\section{Xilinx NetFPGA based prototype testbed}

This section presents the details of the proof-of-concept prototype
implementation of the proposed SMARTHO architecture using Xilinx
NetFPGAs. The implementation details of the testbed, its architecture,
and evaluation results, challenges faced in the development and the
performance results of the testbed is also discussed.

\begin{figure}[htbp]

\begin{subfigure}[b]{\textwidth}
\centering
  \includegraphics[width=0.3\textwidth]{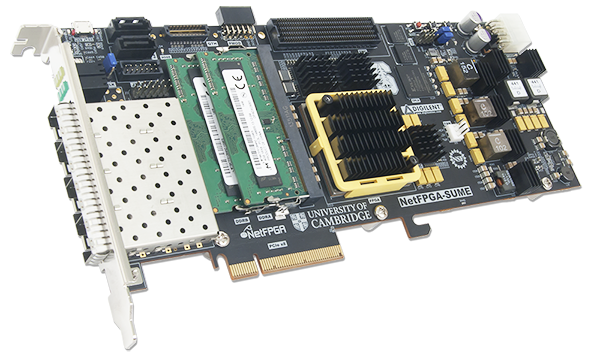}
  \caption{NetFPGA SUME Board}
  \label{fig:netfpgasume}
\end{subfigure}

\begin{subfigure}[b]{\textwidth}
  \includegraphics[width=0.8\textwidth]{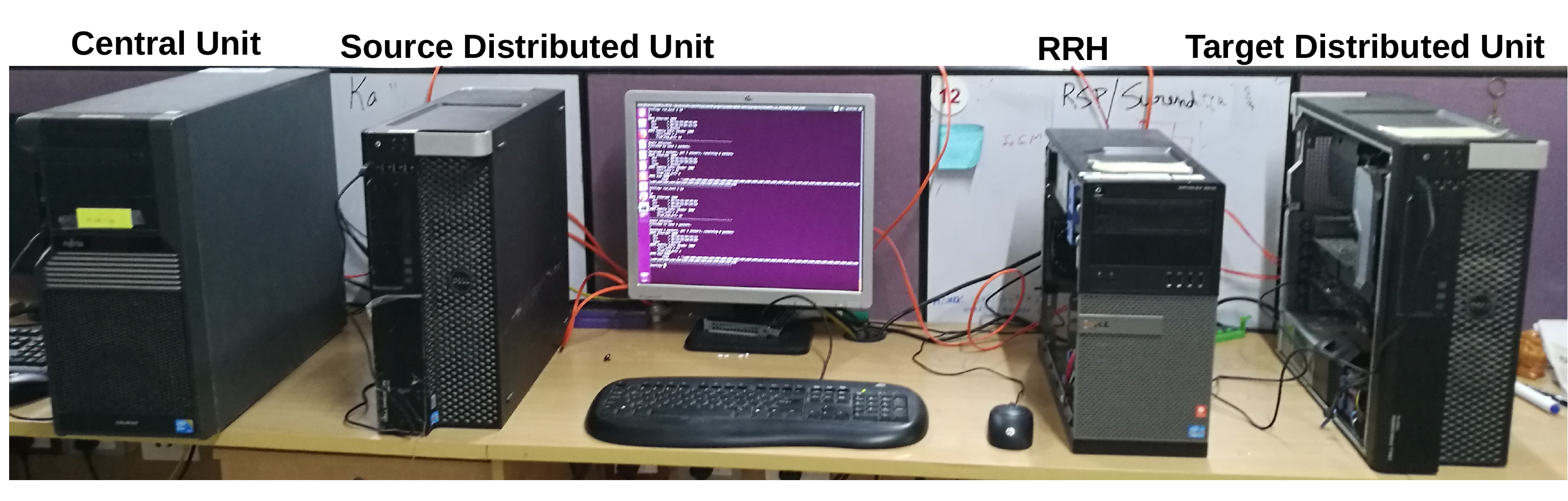}
\centering
  \caption{Overall Prototype View}
  \label{sysconnection}
\end{subfigure}

\begin{subfigure}[b]{0.4\textwidth}
\centering
  \includegraphics[width=0.4\textwidth]{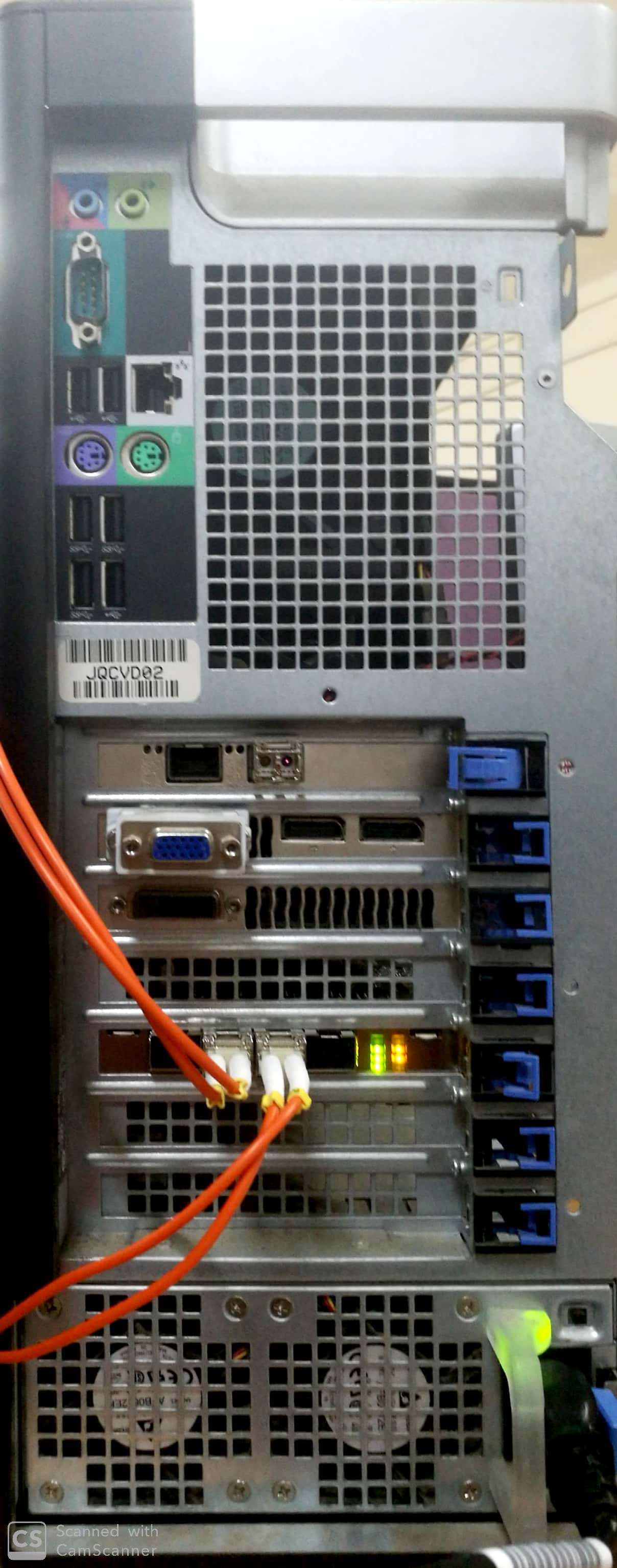}
  \caption{Ports nf1 and nf2.}
  \label{sysconnection2}
\end{subfigure}
\begin{subfigure}[b]{0.55\textwidth}        
\centering
  \includegraphics[width=\textwidth]{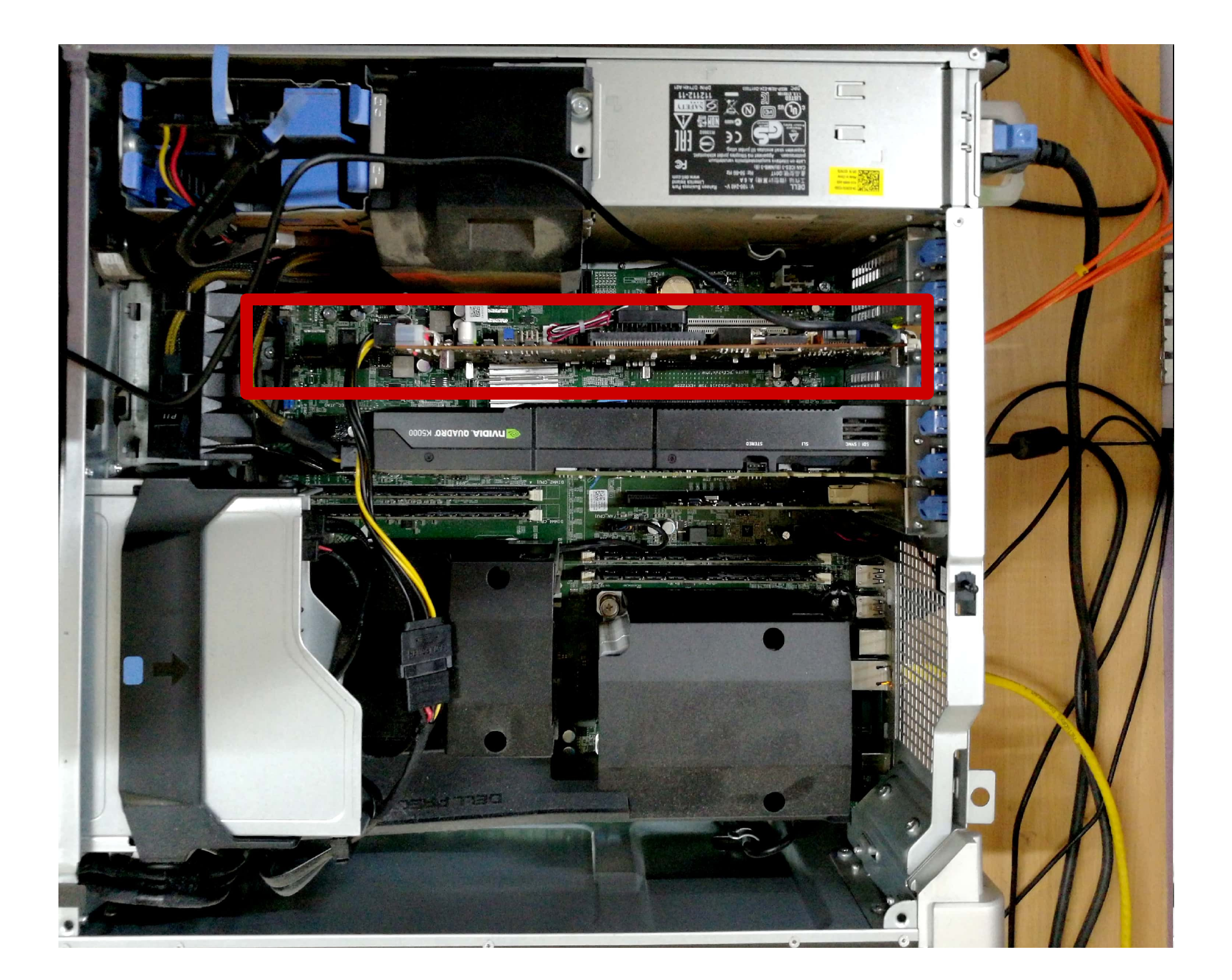}
  \caption{NetFPGA SUME with PCIe connection.}
  \label{fig:netfpgasumesyscon}
\end{subfigure}
\caption{Testbed setup for evaluating SMARTHO performance.}
\end{figure}

\subsection{Xilinx NetFPGA-SUME based prototype}

Xilinx NetFPGA-SUME boards \cite{zilberman2014netfpga} were used as P4
switches for the testbed implementation. The NetFPGA-SUME boards
(Figure~\ref{fig:netfpgasume}) enable researchers to prototype
high-performance applications in hardware. We used Xilinx SDNet
toolchain \cite{netfpgap4}, which simplifies the design of packet
processing data planes that target FPGA hardware. The overall
prototype system is shown in Figure~\ref{sysconnection}.

Four hosts are needed to emulate the behavior of Intra\_CU\_HO, shown
earlier in Figure~\ref{simarch}. The testbed set-up has three
Intel-Xeon, 2.6 GHz i7 core CPU with 64 GB RAM for Source\_DU,
Target\_DU, and CU. For RRH we used Intel Core i7, 32GB RAM. Systems
are integrated with 10G Ethernet and NetFPGA-SUME switch, as shown in
Figure~\ref{sysconnection}. SFP+ fiber-optic LC connector ports are
fixed to NetFPGA-SUME and 10G Ethernet boards, and boards are
connected with LC 50/125 optical fibers as shown in
Figure~\ref{sysconnection2}.

\begin{figure}[htbp]
	\begin{center}
		\includegraphics[width=\textwidth]{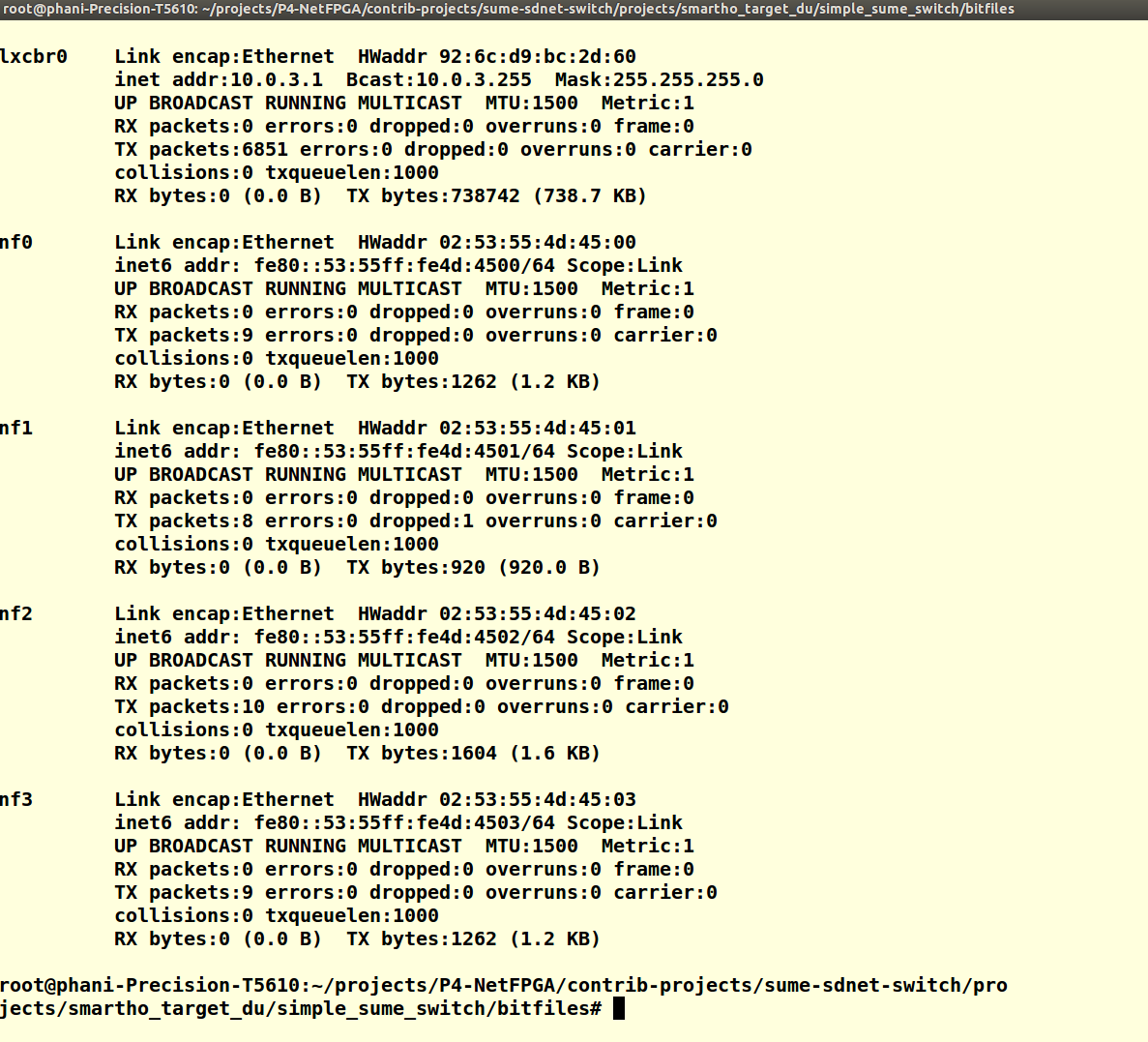}
        \caption{Four ports of NetFPGA after installation of PCI-e drivers.} 
             \label{fig:portsofnetfpgasume}
	\end{center}
\end{figure}

The NetFPGA-SUME board is installed in the host PCI-e slot, as shown
in Figure~\ref{fig:netfpgasumesyscon}. NetFPGA\_Sume boards have four
SFP+ 10Gbps ports, Xilinx refers to these interfaces as nf0, nf1, nf2,
and nf3 where nf0 is the port closest to the link lights on the
board. Loading driver modules (summe\_riffa), the ports on
NetFPGA-SUME is recognized, as shown in
Figure~\ref{fig:portsofnetfpgasume}. These network interfaces are the
means by which the host machine can communicate with the dataplane in
the FPGA.

For traffic generation, a 10Gbps Ethernet card is used at the RRH. The
systems are interconnected using nf1 and nf2 ports. Listed below are
the port connections for the testbed. 

\begin{itemize}
	\item nf1 port of RRH is connected to nf1 port of Source\_DU 
	\item nf2 port of Source\_DU connected to nf1 port of Target\_DU
	\item nf2 port of Target\_DU is connected to nf1 port of CU
\end{itemize}
Overall, three optical LC 50/125 optical fibres and six SFP+ fiber-optic LC connector ports are used.

\subsection{Working model}\label{HOtestbed}
For simulating mobile traffic, we used the Scapy tool \cite{scapy} at
the RRH. The Scapy tool generates raw data packets that mimic the
control messages for HO. The Scapy programs are executed at RRH and
custom packets are created to emulate the control messages of
Intra\_CU\_HO. The format of the header is shown in
Procedure~\ref{smarthoheader}.  These custom packets defined in Scapy
were used as mobile HO control messages. By invoking Smartho(2,4) a
packet is created with control\_information as 2 and forwarding\_tag
as 4. The forwarding\_tag field is used to set the destination port,
control\_information will represent a HO message as shown in
Figure~\ref{contrlmessages}, i.e., Smartho(1,4) represent the
measurement report, Smartho(2,4) represents the uplink RRC measurement
message, and so on.
\floatname{algorithm}{Procedure}
\begin{algorithm}[htbp]
  \caption{smartho\_header.py}\label{smarthoheader}
  \begin{algorithmic}[1]
  \Statex \textbf{class}\text{ Smartho(Packet)}
  \Statex \hspace{1cm} \text{name =}$"Smartho"$,
  \Statex \hspace{1cm} \text{fields\_desc = [}
  \Statex \hspace{2cm} \textit{IntField}\text{("ctrl\_info",1),}
  \Statex \hspace{2cm} \textit{IntField}\text{("frwd\_tag\_prt",4),}
  \Statex \hspace{1cm} \text{]}
  \Statex \hspace{1cm} \textit{def}\text{mysummary(self):}
  \Statex \hspace{2cm} \text{return }\text{self.sprintf("ctrl\_info=\%ctrl\_info\% frwd\_tag\_prt=\%frwd\_tag\_prt\%")} 
  \Statex \text{bind\_layers(Ether, Calc, type=SMARTHO\_TYPE)}
  \Statex \text{bind\_layers(Smartho, Raw)}
  \end{algorithmic}
\end{algorithm}

The Xilinx SDNet tool provides the metadata list to configure the
destination port and to know the source port. There is also one bit
for each of the interfaces (nf0, nf1, nf2, and nf3) in the src\_port
and dst\_port fields (bits 1, 3, 5, and 7). So for example, if the
data-plane wants to send a packet up to the host and have it arrive on
the nf0 Linux network interface then it must set bit 1 of the
dst\_port field (e.g., dst\_port = 0b00000010). The destination port
can be set by the P4 program with variable
sume\_metadata.dst\_port. When sume\_metadata.dst\_port is set as one
the packets are egress to nf0 port, for SMARTHO, all the systems are
connected using nf1 and nf2 ports alone i.e., sume\_metadata.dst\_port
should be either set to 4 or 16.

There are three operations performed by the NetFPGA-SUME switches in
SMARTHO testbed implementation:

\begin{description}
\item \textbf{Change of control information:} 
Control message received at the host (discussed in
Section~\ref{HOtestbed}), is changed as per the next sequence message,
as shown in Figure~\ref{contrlmessages} \&
Figure~\ref{contrlmessages2}.

\item \textbf{Setting the sume\_metadata.dst\_port:}
Extract the value of forwarding\_tag\_port and set this to
sume\_metadata.dst\_port. This would set the egress port for the
control message.

\item \textbf{Change the forwarding\_tag\_port:}
We used look\_up\_table to change the forwarding\_tag\_port. The
look\_up\_table is statically set and does an exact match with control
message and src\_port information put together. Since the architecture
setup is static, the look\_up\_table is loaded at compile time. Based
on the control\_message and sume\_metadata.src\_port the forwarding
port is decided. This forwarding information will be extracted and
updated to the header forwarding\_tag\_port.  The
forwarding\_tag\_port will then be used to set the egress port by the
next host. 

\end{description}

\subsubsection{Traditional HO process}
\begin{figure}[p]
	\begin{center}
		\includegraphics[width=0.85\textwidth]{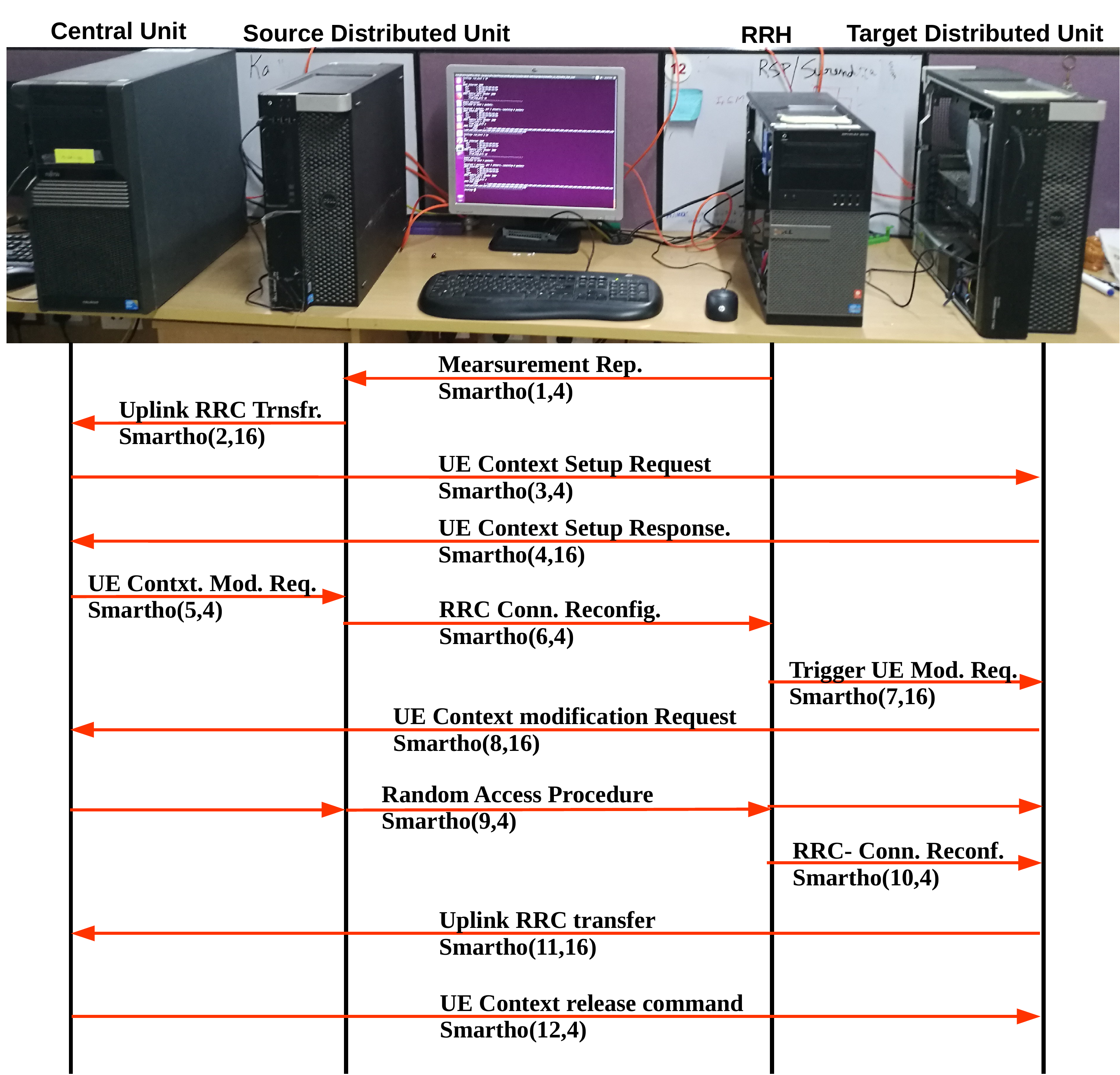}
        \caption{Control message simulating Traditional HO approach on testbed.} 
             \label{contrlmessages}
	\end{center}
\end{figure}

For traditional HO, the sequence of messages that is exchanged in the
testbed is shown in Figure~\ref{contrlmessages}. This emulation
represents the complete traditional HO procedure. The custom function
Smartho(x,y) would add a custom header over Ethernet
(Procedure~\ref{smarthoheader} shows the high-level packet
contents). The initial control message (Measurement report)
Smartho(1,4) is sent from RRH, to Source\_DU to trigger the HO
process.

In the Figure~\ref{contrlmessages}, the exchange of sequence of HO
messages is shown. In the preparation phase the Measurement Report
(MR) message from the RRU would be transmitted to the CU. CU would
select the Target\_DU for the HO by sending HO request (UE Context
Request - Smartho(3,4)). In reality, the UE Context Request should
contain Target-DU-ID, UE context information \& UE History
Information. In testbed we emulated the process creating a custom
header with context information as three. Upon Target\_DU receiving
the HO request it begins handover preparation to ensure seamless
service provision for the UE. The Target\_DU would respond with
setting up Access Stratum (AS) security keys, uplink bearers
connecting to the backhaul, reserve Radio Resource Control (RRC)
resources to be used by the mobile device over the radio link and
allocates Cell-Radio Network Temporary Identifier. This was not
implemented in Target\_DU hosts. To emulate this, we add a 2~ms
delay after sending a Smartho(3,4) message. HO execution and
completion phase are emulated as data exchange as shown in
Figure~\ref{contrlmessages} between the hosts. 
\begin{figure}[p]
		\centering
        \includegraphics[width=0.9\textwidth]{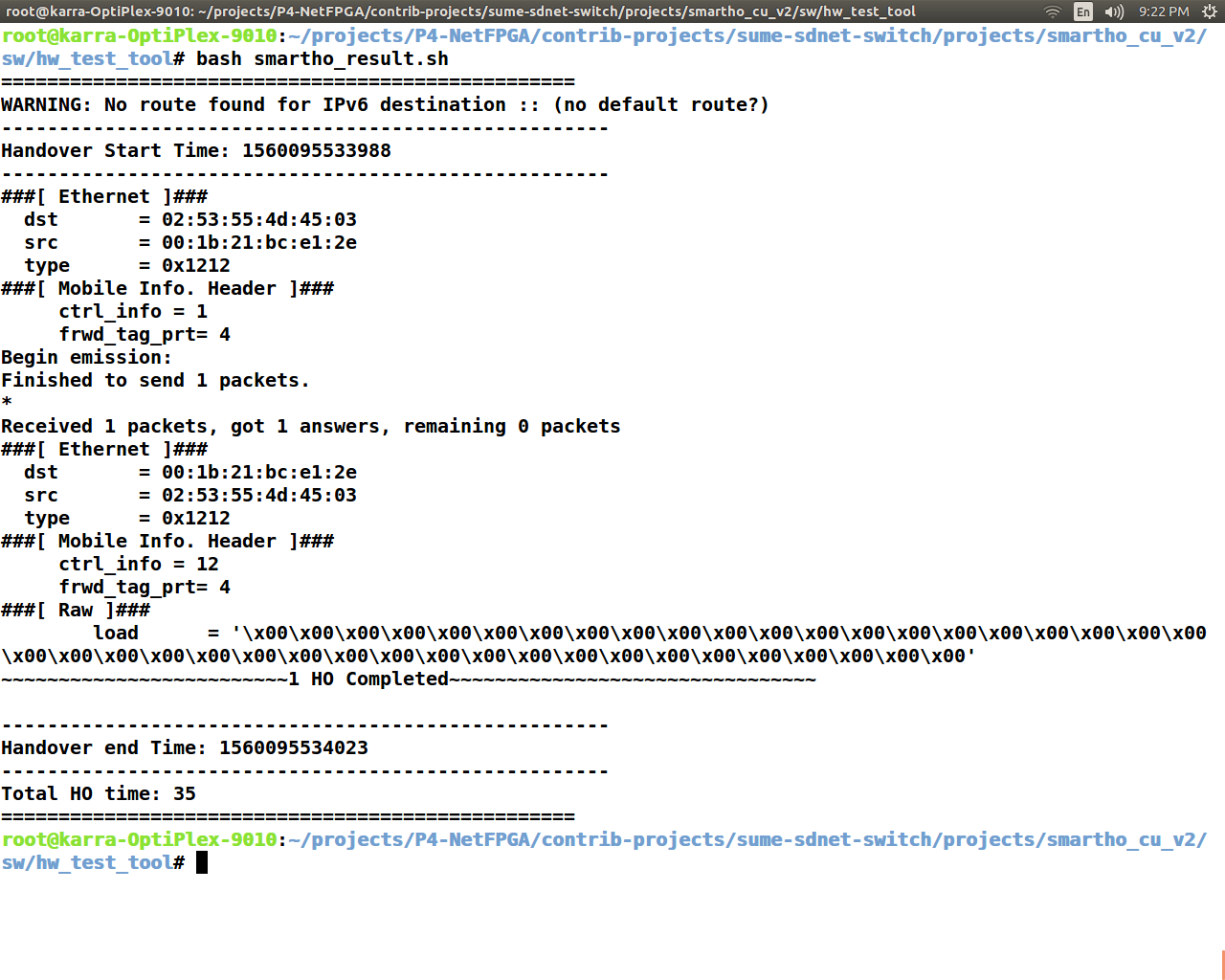}
        \caption{Screenshot showing the results of traditional HO time, with one HO.}
        \label{trad_1}
\end{figure}
\begin{figure}[p]
		\centering
        \includegraphics[width=0.9\textwidth]{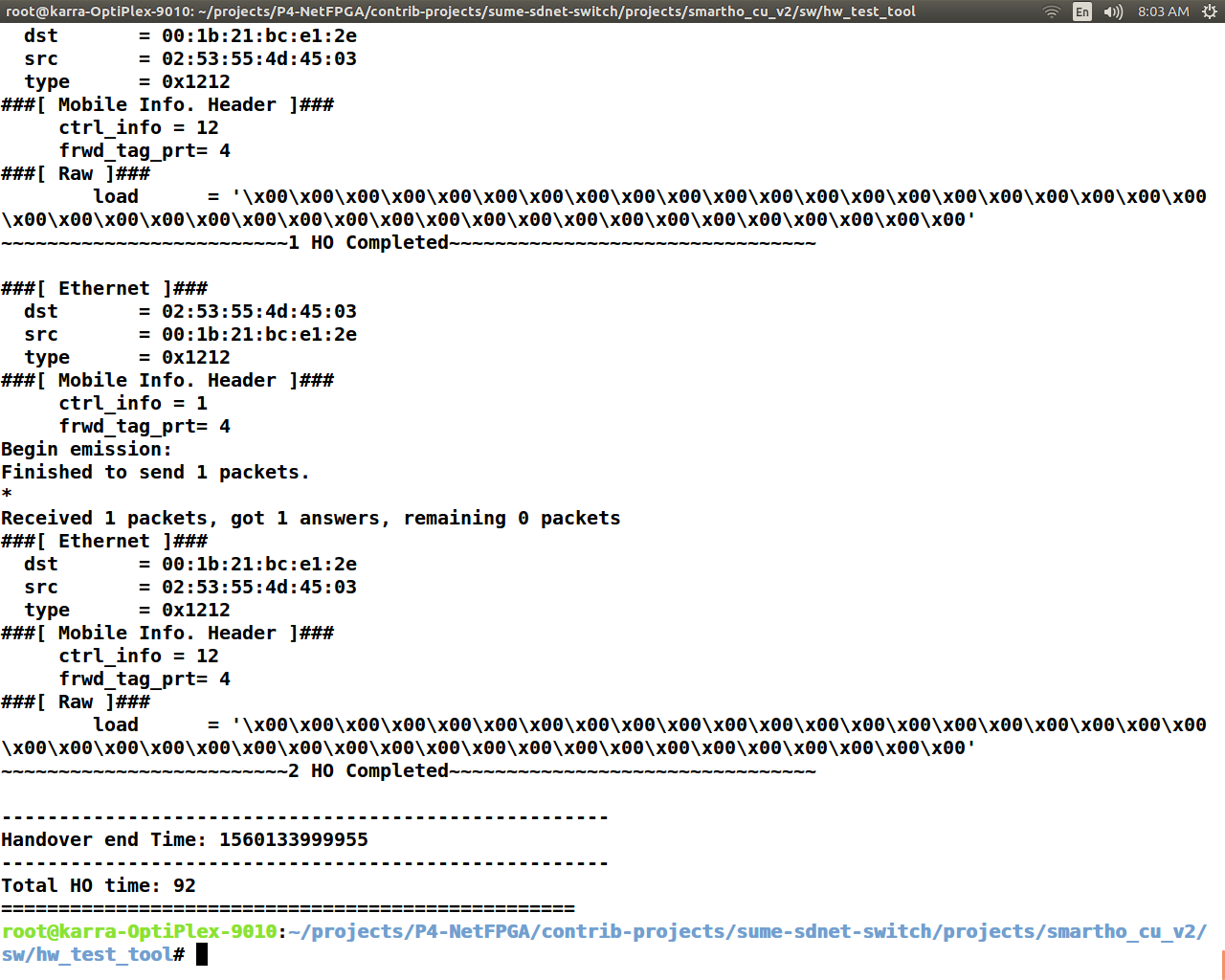}
        \caption{Screenshot showing the results of traditional HO time, with two HOs in tandem.}
        \label{trad_2}
\end{figure}

For simulation of more than one tandem of HOs we sent Smartho(1,4)
multiple times from RRH, i.e., after the first HO is completed, RRH
would resend Smartho(1,4) to Source\_DU to simulate the subsequent
HO. Figure~\ref{trad_1} \& Figure~\ref{trad_2} shows the screenshot
of traditional HO with single and two HOs in tandem.

\subsubsection{SMARTHO process}
For a single HO, the SMARTHO process does not show any difference from
the traditional HO approach. This was discussed in
Section~\ref{smarthoimpl}.  

\begin{figure}[p]
    \centering
	\begin{center}
		\includegraphics[width=0.9\textwidth]{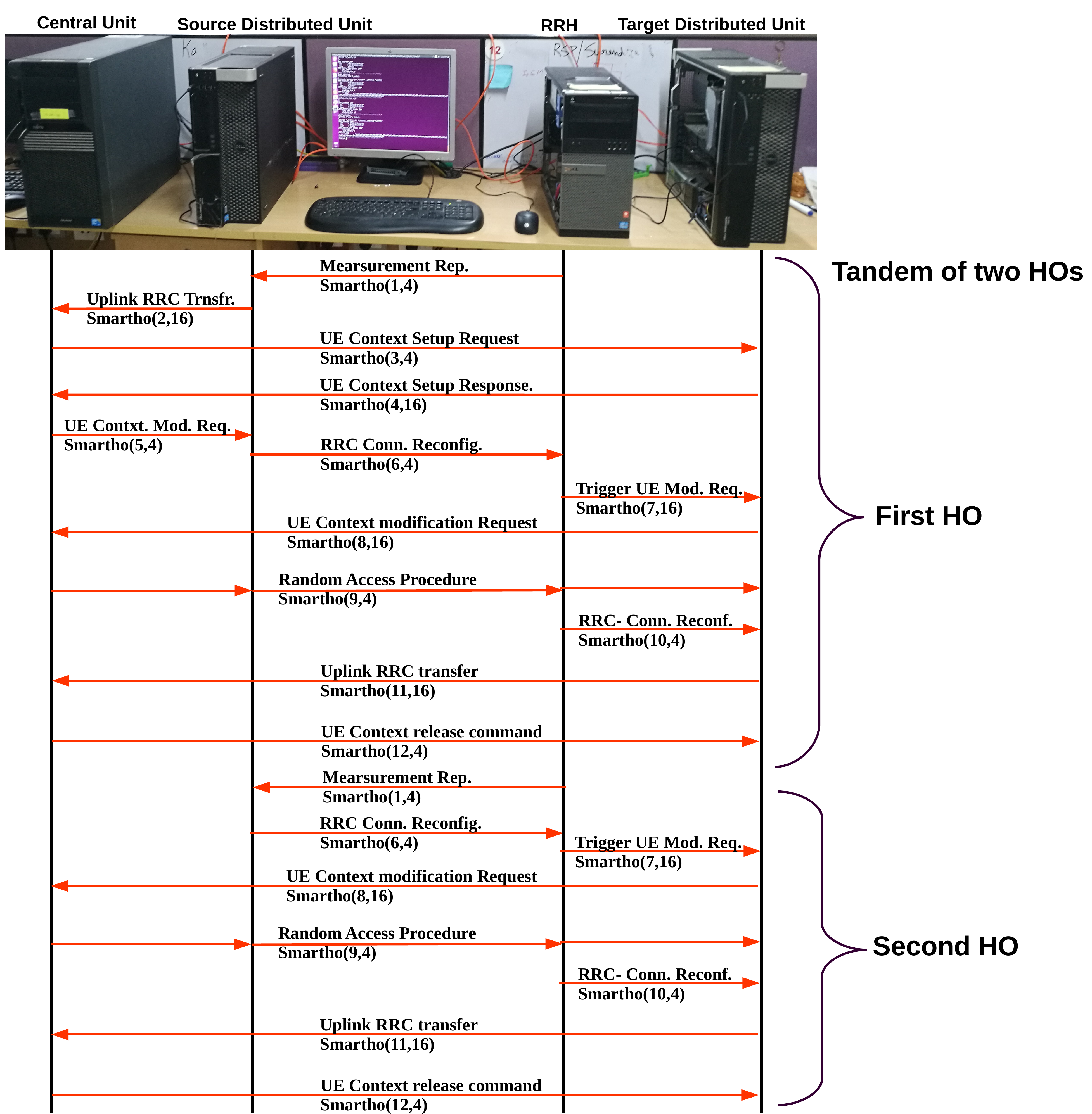}
        \caption{Control message simulating SMARTHO HO behaviour on testbed.} 
             \label{contrlmessages2}
	\end{center}
\end{figure}

\begin{figure}[p]
		\centering
        \includegraphics[width=0.9\textwidth]{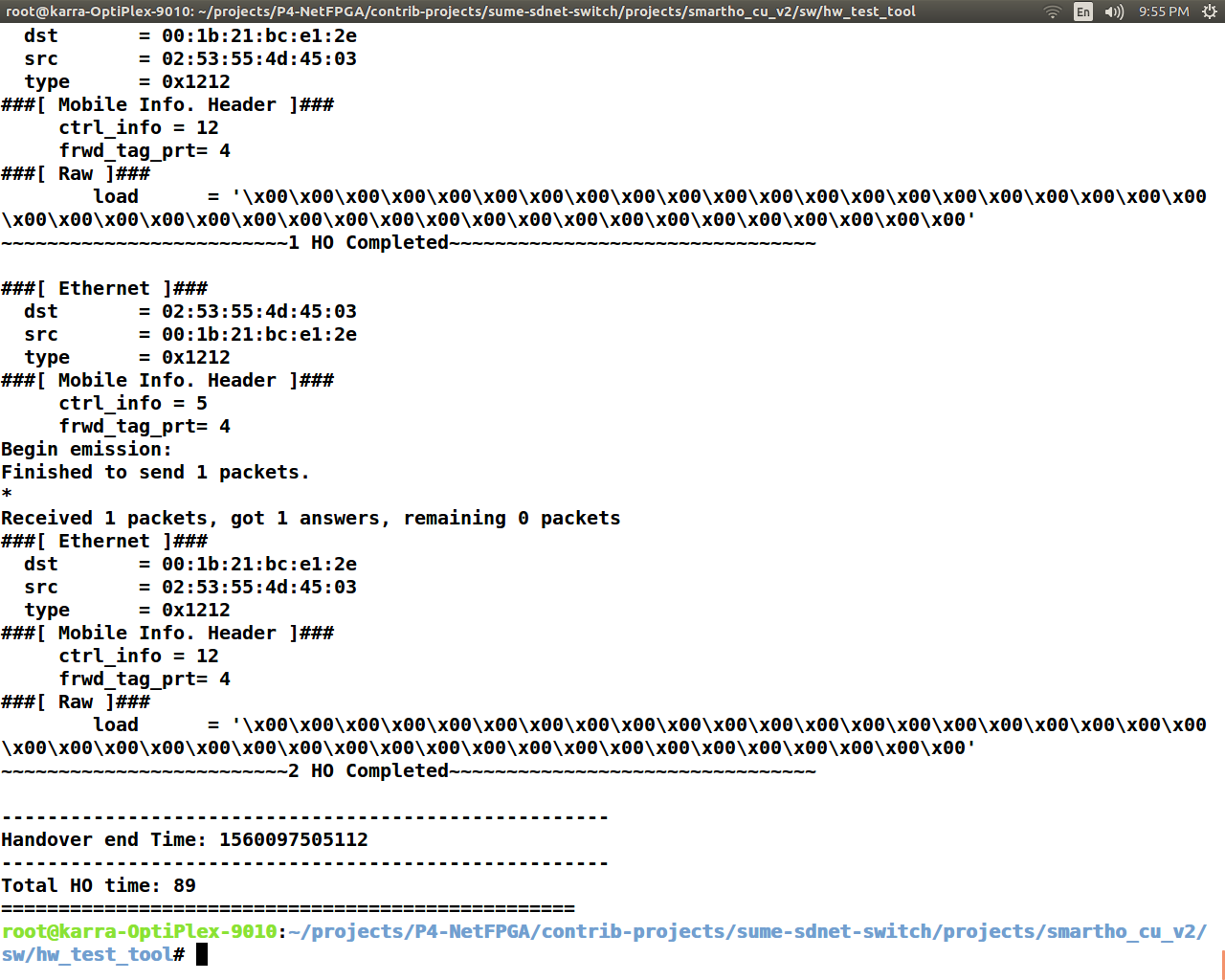}
        \caption{Screenshot showing the results of HO time with
          proposed SMARTHO approach, with two HOs in tandem.} 
        \label{smartho_2}
\end{figure}

\begin{figure}[p]
		\centering
       \includegraphics[width=0.8\textwidth]{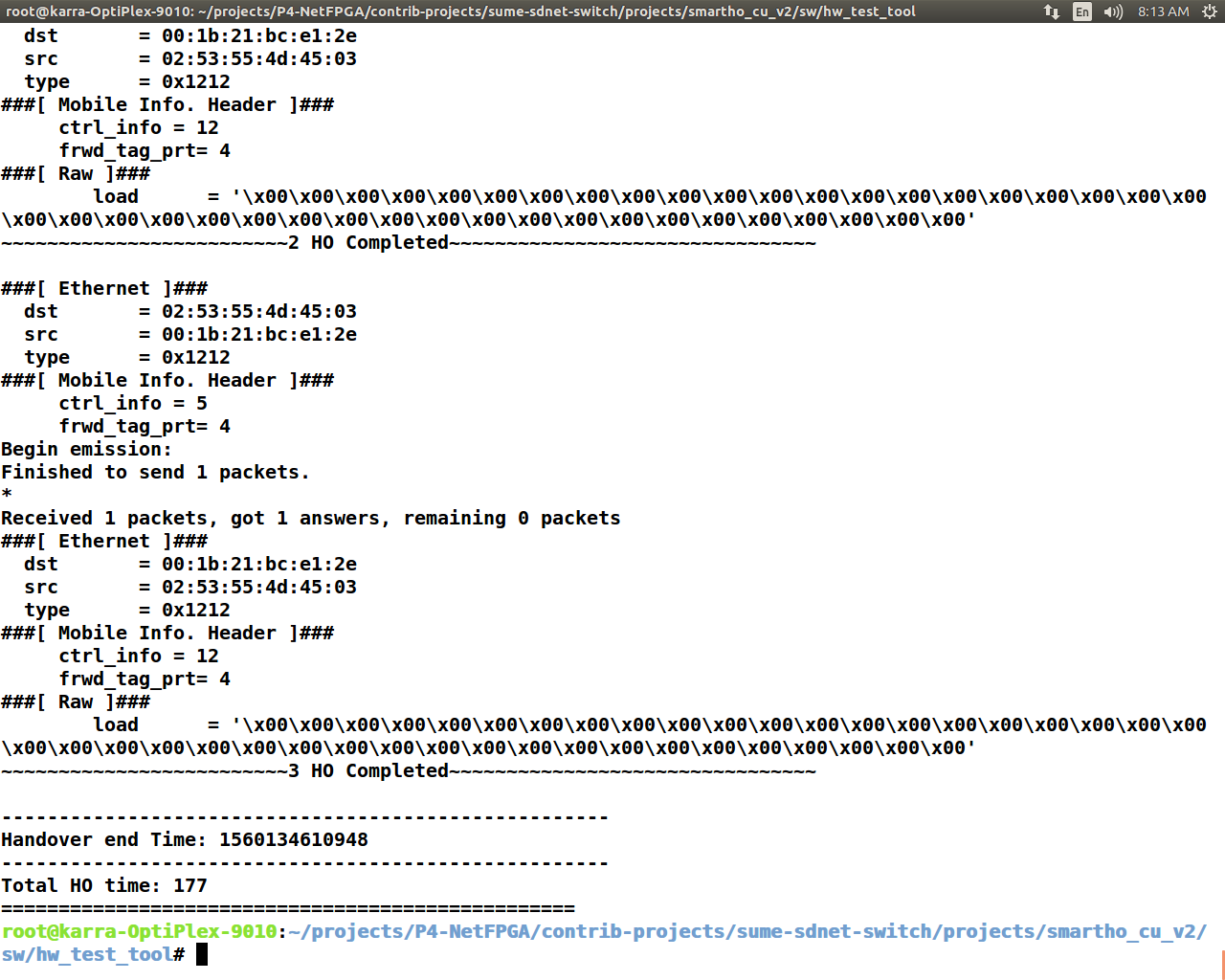}
        \caption{Screenshot showing the results of HO time with
          proposed SMARTHO approach, with two HOs in tandem.} 
       \label{smartho_3}
\end{figure}

The control messaging for SMARTHO process with the tandem of two HOs
is shown in Figure~\ref{contrlmessages2}. Here, after first HO is
completed, the subsequent HOs are performed from the HO execution
phase. In the testbed, we did not implement the controller part, i.e.,
after the first HO messages are executed, the subsequent HOs will send
message Smartho(6,4) as a reply for message
Smartho(1,4). Figure~\ref{smartho_2} shows the screenshot of SMARTHO
for two-HOs in tandem; similar screenshot is shown in 
Figure~\ref{smartho_3} to demonstrate SMARTHO
for three-HOs in tandem.

\begin{figure}[tbp]

\begin{subfigure}[b]{\textwidth}
  \centering
  \includegraphics[width=0.7\textwidth]{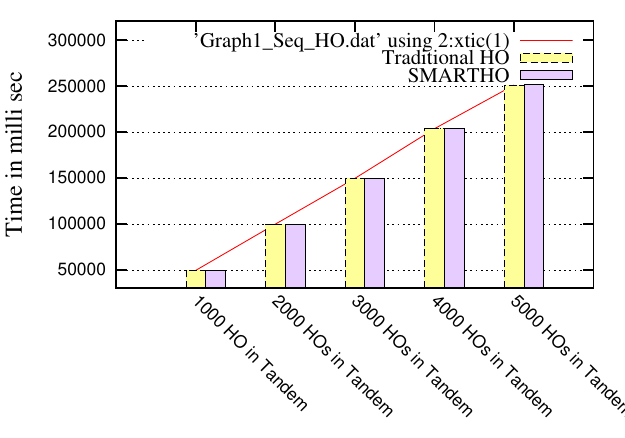}
  \caption{Performance of HO in SMARTHO and Traditional approach.} 
  \label{testbedperformance}
\end{subfigure}

\begin{subfigure}[b]{\textwidth}
  \centering\renewcommand\arraystretch{1.25}
  \begin{tabular}{P{1.0in}*{8}{P{1.0in}}}
\hline \\ \hline
    & \multicolumn{2}{c}{Total HO Time (ms)} & \multicolumn{2}{c}{Time per HO (ms)} \\
    \cmidrule[0.6pt](lr){2-5}\cmidrule[0.6pt](lr){6-9}
    Number of HOs in Tandem & Traditional Approach & SMARTHO & Traditional Approach & SMARTHO \\
    \midrule[\heavyrulewidth]
    1,000 & 49,556 & 49,393 & 49.556 & 49.393 \\
    2,000 & 99,824 & 99,356 & 49.912 & 49.678 \\
    3,000 & 149,693 & 149,305 & 49.897 & 49.768 \\
    4,000 & 213,782 & 203,911 & 53.445 & 50.977 \\
    5,000 & 251,660 & 251,264 & 50.332 & 50.252 \\
    \bottomrule
  \end{tabular}
  \caption{Performance results varying the total number of handovers, executed in tandem.}
  \label{tab:perf}
\end{subfigure}
\caption{Performance results from the prototype implementation.}
\label{perfres}
\end{figure}

Figure~\ref{perfres} presents the performance of SMARTHO and the
Traditional HO approaches, obtained using the prototype. The
experiments are performed on a single UE. In future work, we can
increase the intensity of HOs by sending HO requesting messages in
parallel, from multiple UEs, to emulate the behavior of LTE-R
nodes. We see that as the number of HOs increases, the total time
taking to complete all the HOs is also increased. It is seen that
handover in both approaches takes around 50 milliseconds. Since we
measure HO per UE, the difference is not much, and also the `Time per
HO' is almost the same in both Traditional and SMARTHO. This is not
exactly what was expected; however, the knowledge and expertise gained
in implementing this in the P4 environment is significant. In future
work, we will continue to investigate the performance bottlenecks and
identify coding changes to improve the overall delay. These
experiments can also be extended to multiple UEs in future work.

Using the above testbed experiments, we have demonstrated the
feasibility of implementation  of SMARTHO in a P4-based programmable
dataplane switch.  

\section{Conclusions}

In this report, we have presented the use P4-based dataplane switches
to improve handover efficiency, in a wireless network. The proposed
approach has been studied using a Mininet implementation. The
experimental results show that the proposed SMARTHO approach does have
benefits over the traditional handover process.
The handover mechanism was implemented on a 
Xilinx NetFPGA based P4 switch and the system's working was
demonstrated. 

As part of future work, the Tag-based approach can be considered for
supporting network slicing and virtualization techniques. Further
detailed experiments with multiple UEs and varying loads can also be
conducted. 

\medskip

\subsubsection*{Acknowledgments}

We thank Mr. Karthik Karra, Dr. Manikantan Srinivasan and Dr. C.S.
Ganesh (IIT Madras) for sharing their insights and feedback.  This
work was supported by an IIT Madras IRDA-2017 award and by a DST-FIST
grant (SR/FST/ETI-423/2016) from Government of India (2017--2022).

\clearpage

\bibliographystyle{IEEEtran}

\begin{thebibliography}{10}
\providecommand{\url}[1]{#1}
\csname url@samestyle\endcsname
\providecommand{\newblock}{\relax}
\providecommand{\bibinfo}[2]{#2}
\providecommand{\BIBentrySTDinterwordspacing}{\spaceskip=0pt\relax}
\providecommand{\BIBentryALTinterwordstretchfactor}{4}
\providecommand{\BIBentryALTinterwordspacing}{\spaceskip=\fontdimen2\font plus
\BIBentryALTinterwordstretchfactor\fontdimen3\font minus
  \fontdimen4\font\relax}
\providecommand{\BIBforeignlanguage}[2]{{%
\expandafter\ifx\csname l@#1\endcsname\relax
\typeout{** WARNING: IEEEtran.bst: No hyphenation pattern has been}%
\typeout{** loaded for the language `#1'. Using the pattern for}%
\typeout{** the default language instead.}%
\else
\language=\csname l@#1\endcsname
\fi
#2}}
\providecommand{\BIBdecl}{\relax}
\BIBdecl

\bibitem{CNSM18}
P.~Palagummi and K.~M. Sivalingam, ``{SMARTHO:} {A} network initiated handover
  in {NG-RAN} using p4-based switches,'' in \emph{Proc. of International
  Conference on Network and Service Management ({CNSM})}, Rome, Italy, Nov.
  2018, pp. 338--342.

\bibitem{PhaniThesis}
P.~Palagummi, ``{Network-Initiated Handover Mechanism for 5G Networks using
  P4-based Programmable Data Plane Switches},'' Master's thesis, Indian
  Institute of Technology, Madras, 2019.

\bibitem{3gpp38401}
3GPP, ``{NG-RAN;Architecture description},'' {3rd Generation Partnership
  Project (3GPP)}, Technical Specification (TS) 38.401, 06 2018, version
  15.1.0.

\bibitem{giannoulakis2014applications}
I.~Giannoulakis, E.~Kafetzakis, G.~Xylouris, G.~Gardikis, and A.~Kourtis, ``{On
  the applications of efficient NFV management towards 5G networking},'' in
  \emph{Proc. of Intl. Conf. on 5G for Ubiquitous Connectivity (5GU)}, 2014,
  pp. 1--5.

\bibitem{hawilo2014nfv}
H.~Hawilo, A.~Shami, M.~Mirahmadi, and R.~Asal, ``{NFV: state of the art,
  challenges, and implementation in next generation mobile networks (vEPC)},''
  \emph{IEEE Network}, vol.~28, no.~6, pp. 18--26, 2014.

\bibitem{abdelwahab2016network}
S.~Abdelwahab, B.~Hamdaoui, M.~Guizani, and T.~Znati, ``{N}etwork {F}unction
  {V}irtualization in 5{G},'' \emph{IEEE Communications Magazine}, vol.~54,
  no.~4, pp. 84--91, 2016.

\bibitem{costa2015sdn}
J.~Costa-Requena, J.~L. Santos, V.~F. Guasch, K.~Ahokas, G.~Premsankar,
  S.~Luukkainen, O.~L. P{\'e}rez, M.~U. Itzazelaia, I.~Ahmad, M.~Liyanage
  \emph{et~al.}, ``{SDN and NFV integration in generalized mobile network
  architecture},'' in \emph{Proc. European Conference on Networks and
  Communications (EuCNC)}, 2015, pp. 154--158.

\bibitem{bosshart2014p4}
P.~Bosshart, D.~Daly, G.~Gibb, M.~Izzard, N.~McKeown, J.~Rexford,
  C.~Schlesinger, D.~Talayco, A.~Vahdat, G.~Varghese \emph{et~al.}, ``{P4:
  Programming protocol-independent packet processors},'' \emph{ACM SIGCOMM
  Computer Communication Review}, vol.~44, no.~3, pp. 87--95, 2014.

\bibitem{netfpgap4}
``{P4 to NetFPGA workflow},''
  \url{https://github.com/NetFPGA/P4-NetFPGA-public/wiki/Workflow-Overview},
  Nov. 2019.

\bibitem{p4}
``{P4 Language Consortium},'' \url{https://p4.org}, Jul. 2020.

\bibitem{3gpp22891}
3GPP, ``{Feasibility Study on New Services and Markets Technology Enablers},''
  {3rd Generation Partnership Project (3GPP)}, Technical Specification (TS)
  22.891, 09 2016, version 14.2.0.

\bibitem{checko2015cloud}
A.~Checko, H.~L. Christiansen, Y.~Yan, L.~Scolari, G.~Kardaras, M.~S. Berger,
  and L.~Dittmann, ``{Cloud RAN for mobile networks: A technology overview},''
  \emph{IEEE Communications Surveys \& Tutorials}, vol.~17, no.~1, pp.
  405--426, 2015.

\bibitem{fluidnet}
K.~Sundaresan, M.~Y. Arslan, S.~Singh, S.~Rangarajan, and S.~V. Krishnamurthy,
  ``Fluidnet: A flexible cloud-based radio access network for small cells,''
  \emph{IEEE/ACM Transactions on Networking}, vol.~24, no.~2, pp. 915--928,
  2016.

\bibitem{thyagaturu2018r}
A.~S. Thyagaturu, Z.~Alharbi, and M.~Reisslein, ``{{R-FFT: Function split at
  IFFT/FFT in unified LTE CRAN and cable access network}},'' \emph{IEEE
  Transactions on Broadcasting}, 2018.

\bibitem{wang2017interplay}
X.~Wang, A.~Alabbasi, and C.~Cavdar, ``{Interplay of energy and bandwidth
  consumption in CRAN with optimal function split},'' in \emph{Proc. of IEEE
  ICC}, 2017, pp. 1--6.

\bibitem{3gpp38801}
3GPP, ``{Study on New Radio Access Technology; Radio Access Architecture and
  Interfaces},'' {3rd Generation Partnership Project (3GPP)}, Technical Report
  (TR) 38.801, 03 2017, version 14.0.0.

\bibitem{3gppcudusplit}
------, ``{Summary of RAN3 status on CU-DU split Option 2 and Option 3, and
  questions/issues for RAN2},'' {3rd Generation Partnership Project (3GPP)},
  Technical document (Tdoc) R2-1700637, 01 2017, r2-1700637.

\bibitem{trainseamless}
L.~Tian, J.~Li, Y.~Huang, J.~Shi, and J.~Zhou, ``Seamless dual-link handover
  scheme in broadband wireless communication systems for high-speed rail,''
  \emph{IEEE Journal on Selected Areas in Communications}, vol.~30, no.~4, pp.
  708--718, 2012.

\bibitem{raildistantenna}
T.~Shuo, K.~Zhao, and H.~Wu, ``Wireless communication for heavy haul railway
  tunnels based on distributed antenna systems,'' in \emph{Proc. IEEE VTC
  (Spring)}, 2016, pp. 1--5.

\bibitem{multitunnelmobility}
J.-T. Park and S.-M. Chun, ``Fast local mobility management with multiple
  tunnel support in heterogeneous wireless networks,'' in \emph{Proc. ACM
  international workshop on Mobility management and wireless access}, 2010, pp.
  45--51.

\bibitem{li2016mobility}
H.~Li and D.~Hu, ``{Mobility prediction based seamless RAN-cache handover in
  HetNet},'' in \emph{Proc. IEEE WCNC}, 2016, pp. 1--7.

\bibitem{seamlessHOLTEWIFI}
A.~A. Mansour, N.~Enneya, and M.~Ouadou, ``{A Seamless Handover Based
  MIH-Assisted PMIPV6 in Heterogeneous Network (LTE-WIFI)},'' in \emph{Proc.
  Intl. Conf. on Big Data, Cloud and Applications}, 2017, p.~67.

\bibitem{mihprotocol}
H.~Mzoughi, F.~Zarai, M.~S. Obaidat, and L.~Kamoun, ``{3GPP LTE-advanced
  congestion control based on MIH protocol},'' \emph{IEEE Systems Journal},
  2015.

\bibitem{lteperformanceonltevel}
R.~Merz, D.~Wenger, D.~Scanferla, and S.~Mauron, ``{Performance of LTE in a
  high-velocity environment: A measurement study},'' in \emph{Proc. Workshop on
  All things cellular: operations, applications, \& challenges}, 2014, pp.
  47--52.

\bibitem{mobperfhetnets}
M.~Mehta, N.~Akhtar, and A.~Karandikar, ``{Impact of handover parameters on
  mobility performance in LTE HetNets},'' in \emph{Proc. NCC}, 2015, pp. 1--6.

\bibitem{handoverinmobility}
K.~Kitagawa, T.~Komine, T.~Yamamoto, and S.~Konishi, ``{A handover optimization
  algorithm with mobility robustness for LTE systems},'' in \emph{{Personal
  Indoor and Mobile Radio Communications (PIMRC)}}.\hskip 1em plus 0.5em minus
  0.4em\relax IEEE, 2011.

\bibitem{lterailtriggeropt}
W.~Luo, X.~Fang, M.~Cheng, and X.~Zhou, ``{An optimized handover trigger scheme
  in LTE systems for high-speed railway},'' in \emph{Proc. Intl. Workshop on
  Signal Design and its Applications in Communications (IWSDA)}, 2011, pp.
  193--196.

\bibitem{zheng2008performance}
N.~Zheng and J.~Wigard, ``{On the performance of integrator handover algorithm
  in LTE networks},'' in \emph{Proc. IEEE VTC (Fall)}, 2008, pp. 1--5.

\bibitem{anas2007performance}
M.~Anas, F.~D. Calabrese, P.~E. Mogensen, C.~Rosa, and K.~I. Pedersen,
  ``{Performance evaluation of received signal strength based hard handover for
  UTRAN LTE},'' in \emph{Proc. IEEE VTC}, 2007, pp. 1046--1050.

\bibitem{martinez2015next}
C.~Martinez, R.~Ferro, and W.~Ruiz, ``{N}ext {G}eneration {N}etworks under the
  {SDN} and {OpenFlow} protocol architecture,'' in \emph{Proc. of Workshop on
  Engineering Applications-International Congress on Engineering (WEA)}, 2015,
  pp. 1--7.

\bibitem{pfaff2015design}
B.~Pfaff, J.~Pettit, T.~Koponen, E.~J. Jackson, A.~Zhou, J.~Rajahalme,
  J.~Gross, A.~Wang, J.~Stringer, and P.~Shelar, ``{The Design and
  Implementation of Open vSwitch},'' in \emph{NSDI}, 2015, pp. 117--130.

\bibitem{hommes2017optimising}
S.~Hommes, P.~Valtchev, K.~Blaiech, S.~Hamadi, O.~Cherkaoui \emph{et~al.},
  ``{Optimising packet forwarding in multi-tenant networks using rule
  compilation},'' in \emph{Proc. IEEE Intl. Symposium on Network Computing and
  Applications (NCA)}, 2017, pp. 1--9.

\bibitem{macdavid2017concise}
R.~MacDavid, R.~Birkner, O.~Rottenstreich, A.~Gupta, N.~Feamster, and
  J.~Rexford, ``{Concise encoding of flow attributes in SDN switches},'' in
  \emph{Proc. ACM SOSR}, 2017, pp. 48--60.

\bibitem{chourasia2015sdn}
S.~Chourasia and K.~M. Sivalingam, ``{SDN based Evolved Packet Core
  architecture for efficient user mobility support},'' in \emph{Proc. IEEE
  NetSoft}, 2015, pp. 1--5.

\bibitem{benavcek2017line}
P.~Ben{\'a}{\v{c}}ek, V.~Pu{\v{s}}, J.~Ko{\v{r}}enek, and M.~Kekely, ``{Line
  rate programmable packet processing in 100Gb networks},'' in \emph{Proc.
  Intl. Conf. on Field Programmable Logic and Applications (FPL)}, 2017, pp.
  1--1.

\bibitem{P4Software}
{P4 language Consortium}, ``P4 behaviour model,''
  \url{https://github.com/p4lang/behavioral-model}, 2013.

\bibitem{shahbaz2016pisces}
M.~Shahbaz, S.~Choi, B.~Pfaff, C.~Kim, N.~Feamster, N.~McKeown, and J.~Rexford,
  ``Pisces: A programmable, protocol-independent software switch,'' in
  \emph{Proceedings of the ACM SIGCOMM Conference}, 2016, pp. 525--538.

\bibitem{p416psa}
\emph{P4 16 Portable Switch Architecture (PSA)}, The P4 Language Consortium, 3
  2018, ver. 1.0.0.

\bibitem{P416}
\emph{P4 16 Language Specification}, The P4 Language Consortium, 5 2017, ver.
  1.0.0.

\bibitem{ralfkreherltesignaling}
K.~G. Ralf~Kreher, \emph{LTE SIGNALING, TROUBLESHOOTING AND PERFORMANCE
  MEASUREMENT}.\hskip 1em plus 0.5em minus 0.4em\relax John Wiley \& Sons,
  2016, ch. 2.3, pp. 166--169.

\bibitem{mininet}
M.~Team, ``Mininet,'' \url{mininet.org}, 2020.

\bibitem{scapy}
{SCAPY}, ``{SCAPY - A Python packet crafting tool},''
  \url{https://github.com/secdev/scapy}, 2020.

\bibitem{fayazbakhsh2014enforcing}
S.~K. Fayazbakhsh, L.~Chiang, V.~Sekar, M.~Yu, and J.~C. Mogul, ``Enforcing
  network-wide policies in the presence of dynamic middlebox actions using
  flowtags.'' in \emph{NSDI}, vol.~14, 2014, pp. 533--546.

\bibitem{zilberman2014netfpga}
N.~Zilberman, Y.~Audzevich, G.~A. Covington, and A.~W. Moore, ``{NetFPGA SUME:
  Toward 100 Gbps as research commodity},'' \emph{IEEE Micro}, vol.~34, no.~5,
  pp. 32--41, 2014.

\end{thebibliography}

\begin{IEEEbiography}{Phanindra Palagummi} 
is currently with Microsoft, Hyderabad, India. He received the
M.S. (by Research) degree in Computer Science and Engineering from
Indian Institute of Technology Madras, Chennai, INDIA in 2019; and the
B.Tech. degree in Computer Science and Engineering from Nova College
of Engineering and Technology, affiliated to JNTU, in 2010.  His
research interests include computer networking.
\end{IEEEbiography}

\begin{IEEEbiography}{Krishna M. Sivalingam} 
is an Institute Chair Professor in the
Department of CSE, IIT Madras, Chennai, INDIA, where he was also Head
of the Department from 2016 till 2019. Previously, he was a Professor
in the Dept. of CSEE at University of Maryland, Baltimore County,
Maryland, USA from 2002 until 2007; with the School of EECS at
Washington State University, Pullman, USA from 1997 until 2002; and
with the University of North Carolina Greensboro, USA from 1994 until
1997.  He has also conducted research at Lucent Technologies' Bell
Labs in Murray Hill, NJ, and at AT\&T Labs in Whippany, NJ.  He
received his Ph.D. and M.S.  degrees in Computer Science from State
University of New York at Buffalo in 1994 and 1990 respectively; and
his B.E.  degree in Computer Science and Engineering in 1988 from Anna
University's College of Engineering Guindy, Chennai (Madras), India.
While at SUNY Buffalo, he was a Presidential Fellow from 1988 to 1991.

His research interests include wireless networks, optical wavelength
division multiplexed networks, and performance evaluation. His work
has been supported by several sources including AFOSR, DST India, DOT
India, IBM, NSF, Cisco, Intel, Tata Power Company and Laboratory for
Telecommunication Sciences.  He holds three patents in wireless
networks and has published several research articles including more
than seventy journal publications. He has co-edited a book on Next
Generation Internet Technologies in 2010; on Wireless Sensor Networks
in 2004; on optical WDM networks in 2000 and 2004.  He is serving or
has served as a member of the Editorial Board for journals including
IEEE Networking Letters, ACM Wireless Networks Journal, IEEE
Transactions on Mobile Computing, and Elsevier Optical Switching and
Networking Journal. He has served as Editor-in-Chief of Springer
Photonic Network Communications Journal and EAI Endorsed Transactions
on Future Internet.

He is a Fellow of IEEE, a Fellow of INAE and an ACM Distinguished
Scientist.
\end{IEEEbiography}
\end{document}